\documentclass[aps,showpacs,nofootinbib,preprintnumbers]{revtex4}

\usepackage{graphicx}
\usepackage{graphics}
\usepackage{amssymb}
\usepackage{amsmath}
\newcommand{\nn}{\nonumber}
\newcommand{\nslash}{\kern 0.2 em n\kern -0.50em /}
\newcommand{\kslash}{\kern 0.2 em k\kern -0.45em /}
\newcommand{\pslash}{\kern 0.2 em p\kern -0.50em /}
\newcommand{\Sslash}{\kern 0.2 em S\kern -0.50em /}
\newcommand{\Pslash}{\kern 0.2 em P\kern -0.50em /}
\newcommand{\Rslash}{\kern 0.2 em R\kern -0.50em /}
\newcommand{\open}{{<\kern -0.3 em{\scriptscriptstyle )}}}

\newcommand{\eps}{\epsilon}
\newcommand{\ii}{{\rm i}}

\newcommand{\de}{d}

\newcommand{\tr}{\text{Tr}}

\begin{document}

\title{
Modeling dihadron fragmentation functions
}

\author{Alessandro Bacchetta}
\email{alessandro.bacchetta@desy.de}
\affiliation{Theory Group, Deutsches Elektronen-Synchroton DESY, \\
D-22603 Hamburg, Germany}

\author{Marco Radici}
\email{marco.radici@pv.infn.it}
\affiliation{Dipartimento di Fisica Nucleare e Teorica, Universit\`{a} di Pavia, 
and\\
Istituto Nazionale di Fisica Nucleare, Sezione di Pavia, I-27100 Pavia, Italy}

%%%%%%%%%%%%%%%%%%%%%%%%%%%%%%%%%%%%%%%%%%%%%%%%%%%%%%%%%%%%
\begin{abstract}
We present a model for dihadron fragmentation functions, describing 
the fragmentation of a quark into two unpolarized hadrons. We tune the
parameters  
of our model to the output of the PYTHIA event generator for two-hadron
semi-inclusive production in deep inelastic scattering at HERMES. 
Once the parameters of the model are fixed, 
we make predictions for other unknown fragmentation functions and 
for a single-spin asymmetry in the
azimuthal distribution of $\pi^+ \pi^-$ pairs in semi-inclusive deep inelastic
scattering on a transversely polarized target at HERMES and
COMPASS. Such asymmetry could be
used to measure
the quark transversity distribution function.
\end{abstract}


\preprint{DESY 06-119}

\maketitle

%%%%%%%%%%%%%%%%%%%%%%%%%%%%%%%%%%%%%%%%%%%%%%%%%%%%%%%%%%%%
\section{Introduction}
\label{sec:intro}

Dihadron Fragmentation Functions (DiFF) describe the probability that a quark
hadronizes into two hadrons plus anything else, i.e.\ the process $q\to
H_1\,H_2\,X$. They can 
appear in any process where hadronization is involved, in 
particular in lepton-lepton, lepton-hadron and hadron-hadron collisions 
producing final-state hadrons. They carry information that is not accessible
to single-hadron 
fragmentation functions, but on the other hand they are more complex to
study and to measure.

Unpolarized DiFF were introduced for the first time by Konishi, Ukawa and
Veneziano~\cite{Konishi:1978yx}. Their evolution equations have
been studied in Refs.~\cite{Vendramin:1981te,Sukhatme:1980vs} and more
recently 
reanalyzed in Refs.~\cite{deFlorian:2003cg,Majumder:2004wh,Majumder:2004br}. 
All these studies focused on the probability of producing two hadrons with
energy fractions $z_1$ and $z_2$ by integrating 
over the invariant mass of the produced
pair. 
However, 
%DiFF can depend also on the pair's invariant mass. 
%In fact, 
it is fair to say that the only experimental
information related to unpolarized DiFF consists of invariant mass spectra 
of hadron pairs produced in 
$e^+e^-$ annihilation~\cite{Acton:1992sa,Abreu:1992xx,Buskulic:1995gm}, 
Semi-Inclusive Deep-Inelastic Scattering 
(SIDIS)~\cite{Cohen:1982zg,Aubert:1983un,Arneodo:1986tc} and proton-proton
collisions~\cite{Blobel:1973wr,Aguilar-Benitez:1991yy,Adams:2003cc}. 
Recently, it has been suggested to use DiFF as tools to investigate the 
in-medium effects in heavy-ion
collisions~\cite{Adams:2003cc,Fachini:2004jx,Majumder:2004wh,Majumder:2004br,Majumder:2005vs}.
To address this and other issues, it is necessary 
to improve our knowledge of unpolarized DiFF in vacuum. 

DiFF can be used also for spin studies. In particular, they can act as 
analyzers of the spin of the fragmenting
quark~\cite{Efremov:1992pe,Collins:1994kq,Collins:1994ax,Jaffe:1998hf,Artru:1996zu}
and they can be used to study vector meson
polarization~\cite{Efremov:1982sh,Ji:1994vw,Anselmino:1999cg,Bacchetta:2000jk}.
The definition and properties of all possible DiFF for two unpolarized detected 
hadrons have been presented in Ref.~\cite{Bianconi:1999cd} up to leading twist, 
and in Ref.~\cite{Bacchetta:2003vn} up to subleading twist integrated over 
the transverse component of the center-of-mass (cm) momentum of
the hadron pair. Despite the wealth of observables related to polarized DiFF, 
experimental information is limited~\cite{Abreu:1997wd,Abbiendi:1999bz,Abe:1994bk}. 

At present, the most important application of polarized DiFF appears to be 
the measurement of the quark transversity distribution in the nucleon. This
function, $h_1$, represents 
the probabilistic distribution of transversely polarized
partons inside transversely polarized hadrons, and is a missing
cornerstone to complete the knowledge of the leading-order (spin) structure of the 
nucleon (for a review see Ref.~\cite{Barone:2003fy}). 
%Its peculiar behavior under 
%evolution represents a basic test of QCD in the 
%perturbative domain~\cite{Jaffe:1996zw}. 
Being a chiral-odd function, $h_1$ 
needs to be combined with another chiral-odd soft function. The simplest 
possibility is to consider double-spin asymmetries
in polarized 
Drell-Yan processes~\cite{Ralston:1979ys}. 
This option is under investigation at BNL using high-energy 
polarized proton-proton collisions~\cite{Soffer:2002tf,Bianconi:2005yj}
and could be studied also at 
GSI using polarized proton-antiproton 
collisions~\cite{Anselmino:2002pd,Efremov:2004qs,Bianconi:2004wu,Bianconi:2005bd}.

Another possibility is to measure Single-Spin Asymmetries (SSA) 
in the SIDIS production of a pion on 
transversely polarized targets. Recent data have been released using 
proton~\cite{Airapetian:2004tw,Diefenthaler:2005gx} and 
deuteron~\cite{Alexakhin:2005iw} targets. Their interpretation advocates the
so-called Collins effect~\cite{Collins:1993kk}, by which
a leading-twist contribution to the
cross section appears where $h_1$ is convolved with the Collins function 
$H_1^\perp$, a fragmentation functions that describes the decay probability of a transversely 
polarized quark into a single pion. However, 
extracting $h_1$ from SSA data requires the cross section to depend explicitly
upon
the transverse momentum of the detected pion with respect to the photon
axis~\cite{Boer:1998nt}. 
This fact brings in several 
complications, including the possible overlap of the Collins effect with other 
competing mechanisms and more complicated factorization proofs and evolution 
equations~\cite{Collins:2004nx,Ji:2004wu}.

Semi-inclusive production of two hadrons~\cite{Collins:1994kq,Jaffe:1998hf} offers
an alternative way to access transversity, 
where the chiral-odd partner of transversity is
represented by the DiFF $H_1^{\open}$~\cite{Radici:2001na}, which relates the 
transverse spin of the quark to the azimuthal orientation of the two-hadron plane.
This function is at present unknown. Very recently, the HERMES collaboration has
reported measurements of the asymmetry containing the product
$h_1 H_1^{\open}$~\cite{vanderNat:2005yf}. The COMPASS collaboration has
also presented analogous preliminary results~\cite{Martin:2006}.  
In the meanwhile, the BELLE collaboration is planning to measure the 
fragmentation functions $H_1^{\open}$ in the near
future~\cite{Hasuko:2003ay,Abe:2005zx}. 

In this context, it seems of great importance to devise a way to model DiFF. From 
the theoretical side, this can help understanding 
what are the essential building 
blocks and mechanisms involved in dihadron fragmentation. It can also provide a 
guidance for fits to data and further phenomenological studies. From the 
experimental side, a model could be useful to study the effects of cuts and 
acceptance, to estimate the size of observables in different processes and 
kinematical regimes. Our work is not the first one in this 
direction~\cite{Jaffe:1998hf,Bianconi:1999uc,Radici:2001na}. 
The model presented here is close to the one 
discussed in Ref.~\cite{Radici:2001na}. However, for the first time we are 
able to 
fix the parameters by comparing our unpolarized DiFF $D_1$ with the output of
the 
PYTHIA event generator~\cite{Sjostrand:2000wi} tuned for HERMES~\cite{Liebing:2004us}. 
Then, without introducing extra parameters, 
we make predictions for the polarized DiFF $H_1^{\open}$ and the related SSA
involving the transversity distribution $h_1$.

The paper is organized as follows. In Sec.~\ref{s:dihadron}, we review the basic
formalism of DiFF and of SIDIS cross section for two-hadron production. In
Sec.~\ref{s:model}, we describe our model for the fragmentation of a quark into two
unpolarized hadrons and give analytic results for DiFF calculated in this
model. In Sec.~\ref{s:d1}, we fix the parameters of the model by comparing
it to the output of the PYTHIA event generator tuned for HERMES kinematics. 
In Sec.~\ref{s:results}, we show numerical predictions for the DiFF and 
for the above-mentioned SSA in 
the kinematics explored by the HERMES~\cite{vanderNat:2005yf} and COMPASS 
collaborations~\cite{Martin:2006}. Finally, in
Sec.~\ref{s:conc} we draw some conclusions.

%%%%%%%%%%%%%%%%%%%%%%%%%%%%%%%%%%%%%%%%%%%%%%%%%%%%%%%%%%%%%%%%%%%%%%%%%%%%%
\section{Basics of dihadron fragmentation functions}
\label{s:dihadron}

\begin{figure}
\includegraphics[width=8cm]{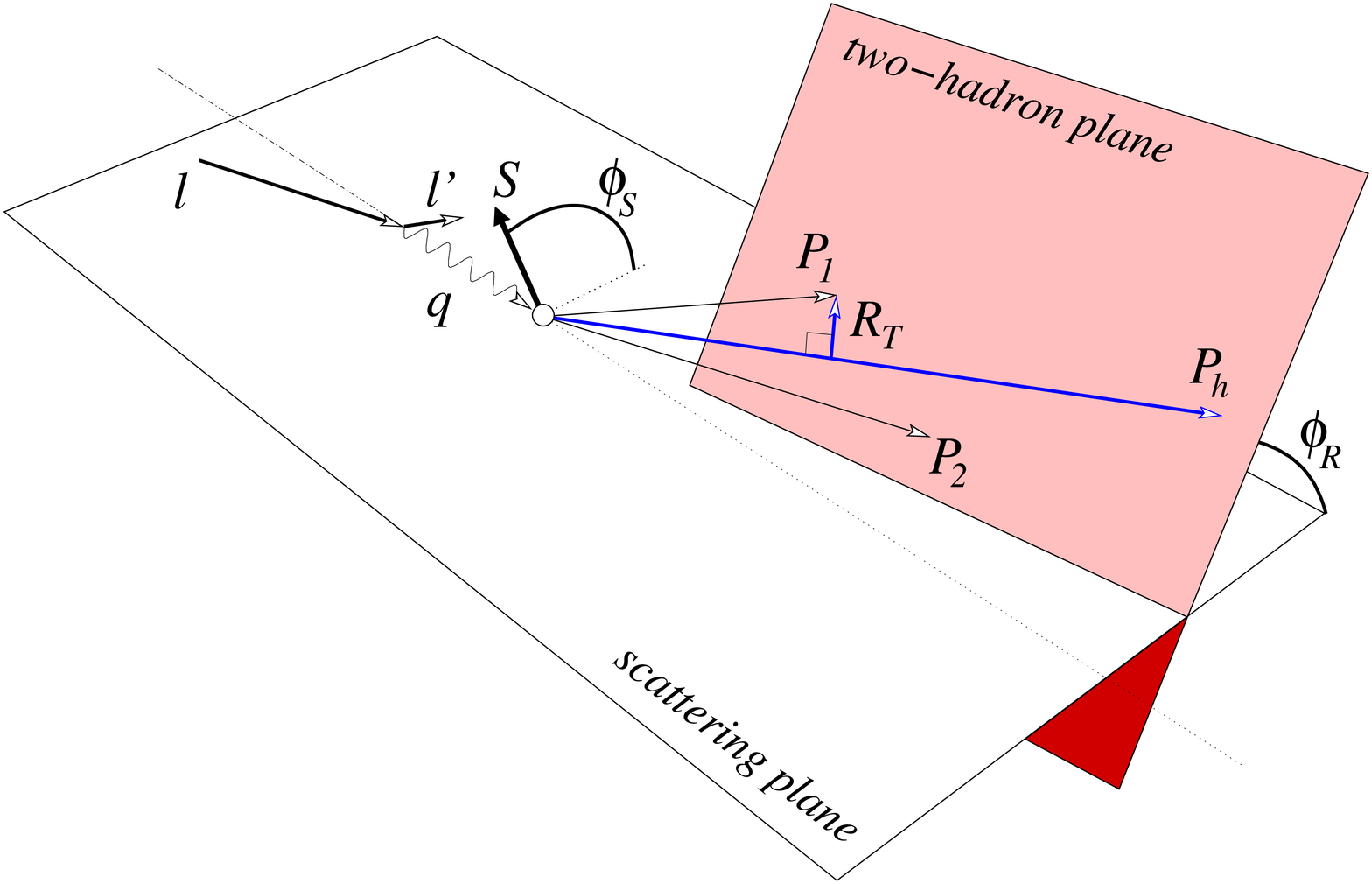}
\caption{Angles involved in the measurement of the transverse  
single-spin asymmetry in deep-inelastic production of two hadrons in the 
current region. 
%The angle $\phi_R$ corresponds to the definition only 
%when  $\pperp \ll Q$.
}
\label{f:sidis} 
\end{figure}

Dihadron Fragmentation Functions are involved in the description of 
the fragmentation process $q\to \pi^+ \pi^- X$. The quark has momentum $k$. 
The two pions have masses 
$m_{\pi} = 0.140$ GeV, momenta
$P_1$ and $P_2$, respectively, and invariant mass $M_h$ 
(considered to be much smaller than the hard scale of the process, e.g., the 
virtuality of the photon, $Q$, in SIDIS).
We introduce the vectors $P_h=P_1+P_2$ and $R=(P_1-P_2)/2$. 
We 
describe a 4-vector $a$ as $[a^-,a^+,a^x,a^y]$, 
i.e.\ in terms of its light-cone 
components $a^\pm = (a^0 \pm a^3)/\sqrt{2}$ and its transverse spatial components. 
We introduce the light-cone fraction $z= P_h^-/k^-$ and 
%$\zeta = 2R^-/P_h^-$ 
the polar angle $\theta$, being the
angle between the direction of $P_1$
in the pair's center of mass and the direction of $P_h$ in the lab 
frame~\cite{Bacchetta:2002ux},
%$\zeta = 2R^-/P_h^-$ 
so that the relevant momenta can be written as
\begin{align} 
   k^\mu &= \biggl[ \frac{P_h^-}{z},\; \frac{z(k^2+\vec k_T^2)}{2P_h^-},\;
        k_T^x, k_T^y 
  \biggr],  
\\
   P_h^\mu &= \biggl[P_h^-,\; \frac{M_h^2}{2 P_h^-},\; 0,\; 0 \biggr],
\\
   R^\mu &= \biggl[\frac{|\vec{R}|\, P_h^-}{M_h}\,\cos{\theta},\;
    - \frac{|\vec{R}|\,M_h}{2\,P_h^-}\,\cos{\theta},\; R_T^x, R_T^y \biggr] \nn 
\\
&= \biggl[\frac{|\vec{R}|\, P_h^-}{M_h}\,\cos{\theta},\;
    - \frac{|\vec{R}|\,M_h}{2\,P_h^-}\,\cos{\theta},\
        |\vec{R}|\,\sin{\theta}\,\cos{\phi_R},\; 
        |\vec{R}|\,\sin{\theta}\,\sin{\phi_R}
\biggr],
\label{e:momenta}
\end{align} 
where 
\footnote{Note that there is a misprint in the expressions for $|\vec{R}|$
  in Eq.~(27) of Ref.~\cite{Bacchetta:2002ux} and in Eq.~(23) of
  Ref.~\cite{Bacchetta:2003vn}.} 
\begin{equation} 
|\vec{R}| = \frac{M_h}{2}\sqrt{1- \frac{4\,m_{\pi}^2}{M_h^2}},
\end{equation} 
and $\phi_R$ is defined later in Eq.~(\ref{angle-def-2}) (see also 
Fig.~\ref{f:sidis}). It is useful to compute the scalar products
\begin{align} 
   P_h \cdot R &= 0,
\\
   P_h \cdot k &= \frac{M_h^2}{2z} + z\, \frac{k^2+|\vec k_T|^2}{2}, 
\\
   R \cdot k &= \biggl(\frac{M_h}{2\,z}
        -z\,\frac{k^2+|\vec{k}_T|^2}{2\,M_h} \biggr)\,
                |\vec{R}|\,\cos{\theta} 
        - \vec{k}_T \cdot \vec{R}_T.
\label{e:scalarproducts}
\end{align} 

Fragmentation functions are extracted from the
correlation function~\cite{Bacchetta:2002ux}
\begin{equation} 
\Delta^q(z,\cos\theta,M_h^2,\phi_R) = 
\frac{z |\vec R|}{16\,M_h}\int d^2 \vec k_T \; 
       d k^+\,\Delta^q(k;P_h,R) \Big|_{k^- = P_h^-/z}  \; , 
\label{eq:delta1}
\end{equation} 
where~\cite{Boer:2003cm,Bacchetta:2003vn}
\begin{equation} \begin{split} 
\Delta^q(k,P_h,R)_{ij}
%&= \frac{1}{4z} \int \de k^+ \;\Delta
%(k,p)\,\bigg|_{k^-=p^-/z} \\
         & =\sum_X \, \int
        \frac{\de^4\xi}{(2\pi)^{4}}\; e^{+\ii k \cdot \xi}
       \langle 0|
%{\cal U}^T{[\infty_T, \vec{\xi}_T;-\infty^+]}\;
%{\cal U}^+{[-\infty^+, \xi^+;\vec{\xi}_T]}
{\cal U}^{n_+}_{(-\infty,\xi)}
\,\psi_i^q(\xi)|P_h, R; X\rangle 
%\\ &\qquad\times
\langle P_h, R;, X|
             \bar{\psi}_j^q(0)\,
%{\cal U}^+{[0^+, -\infty^+;0_T]}\;
%{\cal U}^T{[0_T, \infty_T; -\infty^+]}
{\cal U}^{n_+}_{(0,-\infty)}
|0\rangle \,.    
\label{e:delta2}
\end{split} 
\end{equation} 
%The Wilson lines more explicitly read
%\begin{align} 
%{\cal U}^+_{(-\infty,\xi)} &\equiv 
%{\cal U}^T{[\infty_T, \vec{\xi}_T;-\infty^+]}\;
%{\cal U}^+{[-\infty^+, \xi^+;\vec{\xi}_T]}
%,
%\\
%{\cal U}^+_{(0,-\infty)} &\equiv
%{\cal U}^+{[0^+, -\infty^+;0_T]}\;
%{\cal U}^T{[0_T, \infty_T; -\infty^+]}
%\end{align} 
%The notation ${\cal U}^+{[a^+,b^+;c_T]}$ indicates a Wilson line running along
%the plus direction from $(0^-,a^+,c_T)$ to $(0^-,b^+,c_T)$, while ${\cal
%  U}^T{[a_T,b_T;c^+]}$ indicates a gauge link running along
%the transverse direction from $(0^-,c^+,a_T)$ to $(0^-,c^+,b_T)$. 
Since we are going to perform the integration
over the transverse momentum
$\vec{k}_T$, the Wilson lines ${\cal U}$ 
can be reduced to unity using a light-cone
gauge.  

The only fragmentation functions surviving after $\vec{k}_T$-integration 
are~\cite{Bianconi:1999cd,Bacchetta:2002ux}
\begin{align} 
D_1^q(z,\cos\theta,M_h^2) &= 4\pi\, \tr[\Delta^q(z,\cos\theta,M_h^2,\phi_R)\,
\gamma^-],
\\
\frac{\eps_T^{ij}\,R_{T j}}{M_h}\, H_1^{\open\, q}(z,\cos\theta,M_h^2)
&=4\pi \, \tr[\Delta^q(z,\cos\theta,M_h^2,\phi_R)\,i\,\sigma^{i -}\,\gamma_5].
\end{align} 
These functions can be expanded in the relative partial waves of the pion pair 
system. Truncating the expansion
at the $p$-wave level we obtain~\cite{Bacchetta:2002ux}
\begin{align}
D_1^{q}(z,\cos{\theta},M_h^2) &\approx D_{1,oo}^{q}(z,M_h^2) + 
D_{1,ol}^{q}(z,M_h^2)\, \cos\theta + D_{1,ll}^{q}(z,M_h^2) \, \frac{1}{4}\,
(3\cos^2\theta -1) \; , 
\label{eq:d1pw}
\\
H_1^{\open\,q}(z,\cos{\theta},M_h^2) &\approx  H_{1,ot}^{\open\,q}(z,M_h^2) + 
H_{1,lt}^{\open\,q}(z,M_h^2) \, \cos\theta \; . 
\label{eq:h1angpw}
\end{align} 
The fragmentation function $D_{1,oo}$ can receive
contributions from both $s$ and $p$ waves, but not from the  interference 
between the two, $D_{1,ol}$ and $H_{1,ot}^{\open}$ 
originate from the interference of $s$
and $p$ waves, $D_{1,ll}$ comes from polarized $p$ waves,
and $H_{1,lt}^{\open}$ originates from the interference of two $p$ waves with
different polarization.
%Integrating over $\cos\theta$ 
%we obtain
%\begin{align}
%\int_{-1}^{1} d \cos\theta \;D_1^{q}(z,\cos{\theta},M_h^2) &=
%2\,D_{1,oo}^{q}(z,M_h^2),
%\int_{-1}^{1} d \cos\theta \;H_1^{\open\,q}(z,\cos{\theta},M_h^2)\,\sin\theta&=
%\frac{\pi}{2}\,H_{1,ot}^{\open\,q}(z,M_h^2),
%\end{align} 

Our model can make predictions for 
the above fragmentation functions as well as for 
transverse-momentum-dependent fragmentation functions, which we do
not consider in this Section.
However, we will focus our attention mainly on the functions $D_{1,oo}$ and
$H_{1,ot}^{\open}$ because of their relevance for transversity measurements in
SIDIS~\cite{Collins:1994kq,Jaffe:1998hf,Radici:2001na,Bacchetta:2004mw}. 

Let's consider in fact the SIDIS process $l\,p \to l'\, \pi^+\, \pi^-\, X$,
where $l$ and $l'$ are the momenta of the lepton before and after the
scattering and $q=l-l'$ is the momentum of the virtual photon.
We consider the cross section differential in 
$\de M_h^2$, $\de\phi_R$, $\de z$, $\de x$, $\de y$, $\de\phi_S$,  
where $z$, $x$, $y$ are the usual scaling variables employed in SIDIS, and
the azimuthal angles are defined so that (see Fig.~\ref{f:sidis})\footnote{The 
definition of the angles is consistent with the so-called Trento 
conventions~\cite{Bacchetta:2004jz}.} 
\begin{align}  
  \label{angle-def-1} 
\cos \phi_S &=  
  \frac{(\hat{ q}\times\vec{l})}{|\hat{ q}\times\vec{ l}|} 
  \cdot \frac{(\hat{ q}\times\vec{ S})}{|\hat{ q} 
     \times\vec{ S}|},  
&  
\sin \phi_S &=  
  \frac{(\vec{ l} \times \vec{ S}) \cdot \hat{ q}}{|\hat{ q} 
     \times\vec{ l}|\,|\hat{ q}\times\vec{ S}|} , \\ 
\cos \phi_R &=    
%\frac{(\hat{ P}_h\times\vec{ l})}{|\hat{ P}_h\times\vec{ l}|} 
%  \cdot \frac{(\hat{ P}_h\times\vec{ R})}{|\hat{ P}_h 
%     \times\vec{ R}|} 
%\,\approx\,  
  \frac{(\hat{ q}\times\vec{ l})}{|\hat{ q}\times\vec{ l}|} 
  \cdot \frac{(\hat{ q}\times\vec{ R_T})}{|\hat{ q} 
     \times\vec{ R_T}|},  
%\\ 
& 
\sin \phi_R  
%&= &  
%  \frac{(\vec{ l} \times \vec{ R}) \cdot \hat{ P}_h}{|\hat{ P}_h 
%     \times\vec{ l}|\,|\hat{ P}_h\times\vec{ R}|} 
&=  
  \frac{(\vec{ l} \times \vec{ R_T}) \cdot \hat{ q}}{|\hat{ q} 
     \times\vec{ l}|\,|\hat{ q}\times\vec{ R_T}|} , \label{angle-def-2}
\end{align}  
where $\hat{ q} = \vec{ q}/|\vec{ q}|$ and $\vec{ R_T}$ is the component of 
$R$ perpendicular to $P_h$. 
 
When the target is transversely polarized, 
we can define the following cross section combinations~\footnote{The 
definition of the angles in Eqs.~(\ref{angle-def-1},\ref{angle-def-2}) is consistent 
with the so-called Trento conventions~\cite{Bacchetta:2004jz} and it is the origin 
of the minus sign in Eq.~(\ref{eq:cross0T}) with respect to Eq.~(43) of 
Ref.~\cite{Bacchetta:2002ux} (compare $\phi_R$ and $\phi_S$ in Fig.~\ref{f:sidis} 
with the analogue ones in Fig.~2 of Ref.~\cite{Bacchetta:2002ux}).}
\begin{align}
\de^6\! \sigma^{}_{UU} &= \frac{\de^6\! \sigma^{\uparrow}+\de^6\! 
\sigma^{\downarrow}}{2} = \sum_q \frac{\alpha^2 e_q^2}{\pi\, 
%s\,x\, y^2
y\, 
%(s\,x\,y - M^2)
Q^2
}\,
     \frac{1-y+y^2/2 + y^2\,\gamma^2/4}{1+\gamma^2}\, f_1^q(x)\,  D_{1,oo}^q(z,M_h^2),  
\label{eq:crossOO} 
\\
\de^6\! \sigma^{}_{UT} &=\frac{\de^6\! \sigma^{\uparrow}-\de^6\! \sigma^{\downarrow}}
{2} =-\sum_q \frac{\alpha^2 e_q^2}{4\,
%s\,x\, y^2
y\, 
%(s\,x\,y - M^2)
Q^2
} \,
    \frac{1-y-y^2\,\gamma^2/4}{1+\gamma^2}\, \sin(\phi_R^{} + \phi_S^{})\,
   h_1^q(x)\,\frac{|\vec{R}|}{M_h}\, H_{1,ot}^{\open q}(z,M_h^2), \label{eq:cross0T}
\end{align} 
where $\alpha$ is the fine structure constant, 
%$s$ is the usual Mandelstam variable, 
$\gamma = 2Mx/Q$,
and $M$ is the mass of the target.
These expressions are valid up to leading twist only. Subleading contributions
are described in Ref.~\cite{Bacchetta:2003vn}. In particular, they give rise
to a term proportional to $\cos\phi_R$ in $\de \sigma^{}_{UU}$ and a term
proportional to $\sin\phi_S$ in $\de \sigma^{}_{UT}$. Corrections at order
$\alpha_S$ were partially studied in Ref.~\cite{deFlorian:2003cg}, but further
work is required.

We can define the asymmetry amplitude
\begin{equation} 
A_{UT}^{\sin(\phi_R^{} + \phi_S^{})}(x,y,z,M_h^2) \equiv  
\frac{1}{\sin (\phi_R^{} + \phi_S^{})}\, 
\frac{\de^6\! \sigma^{}_{UT}}{\de^6\! \sigma^{}_{UU}} = -
\frac{\frac{1-y-y^2\,\gamma^2/4}{x\,y^2\,(1+\gamma^2)}}
{\frac{1-y+y^2/2 + y^2\,\gamma^2/4}{x\,y^2\,(1+\gamma^2)}}\,
\frac{\pi\, |\vec{R}|}{4\,M_h} 
\frac{\sum_q e_q^2\,h_1^q(x)\ H_{1,ot}^{\open q}(z,M_h^2)}
{\sum_q e_q^2\,f_1^q(x)\  D_{1,oo}^q(z,M_h^2)}.
\label{e:asy2}
\end{equation} 
Note that we avoided simplifying the prefactors because numerator and
denominator are usually integrated separately over some of the variables.
%If instead of integrating the cross sections 
%over $\cos\theta$, the $\sin \theta$ amplitude is extracted from a fit of the
%unintegrated cross sections, the relative amplitude turns out to be
%\begin{equation} 
%A_{UT}^{\sin(\phi_R^{} + \phi_S^{})\,\sin\theta}(x,y,z,M_h^2) \equiv 
%\frac{4}{\pi}\, A_{UT}^{\sin(\phi_R^{} + \phi_S^{})}(x,y,z,M_h^2).
%\label{e:asy3}
%\end{equation} 

%%%%%%%%%%%%%%%%%%%%%%%%%%%%%%%%%%%%%%%%%%%%%%%%%%%%%%%%%%%%%%%%%%%%%%%%%%%%%
\section{Fragmentation functions in a spectator model}
\label{s:model}

\begin{figure}
\includegraphics[width=8cm]{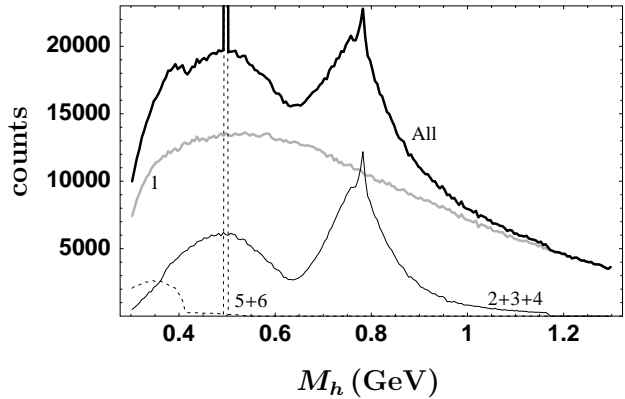}
\caption{Semi-inclusive dihadron counts in bins of $M_h$ 
 from the PYTHIA event generator~\cite{Sjostrand:2000wi} tuned
  for HERMES~\cite{Liebing:2004us}. 
  The thick solid line represents the sum of all
  channels. 
  The thin solid line represents the sum of channels
  2,3, and 4 described in the text. The dashed line represents the sum of
  channels 5 and 6 (which are excluded in our model). The gray line is the
  difference between the total and the sum of all channels 2 to 6 and
  is assumed to represent channel 1.}
\label{f:pythia}
\end{figure}

We aim at describing the process $q\to \pi^+ \pi^- X$ at invariant mass
$M_h\lesssim 1.3$ GeV. 
To have an idea of the prominent channels contributing to this process, we
examined 
the output of
the PYTHIA event generator~\cite{Sjostrand:2000wi} tuned for HERMES~\cite{Liebing:2004us}, 
which well reproduces the measured
events at HERMES. 
Further details concerning the event generator's output will be discussed in the
next section.
Fig.~\ref{f:pythia} shows the number of counted dihadron pairs in bins of
$M_h$ (200 bins from 0.3 to 1.3 GeV).
% and $z$ (200 bins from 0.2 to 0.8).
The total amount of events is 2667889.

A few prominent channels contribute to this process:
\newcounter{saveenum}
\begin{enumerate}
\item{\label{bkg}
$q \to \pi^+ \pi^- X_1$: 
fragmentation into an ``incoherent'' $\pi^+ \pi^-$ pair that we will call, in the 
following, ``background'';}
\item{\label{rho}
$q \to \rho\ X_2 \to \pi^+ \pi^- X_2$: fragmentation into a $\rho$ resonance 
decaying into $\pi^+ \pi^-$, responsible for a peak at $M_h \sim$ 770 MeV
(14.81\%);}
\item{\label{omega2pi}
$q \to \omega\ X_3 \to \pi^+ \pi^- X_3$: fragmentation into a $\omega$ resonance 
decaying into $\pi^+  \pi^-$, responsible for a small peak at $M_h \sim$ 782
MeV (0.31\%);}
\item{\label{omega3pi}
$q \to \omega\ X_4' \to \pi^+ \pi^- X_4$ with $X_4 = \pi^0\ X_4'$: 
fragmentation into a $\omega$ resonance decaying into $\pi^+  \pi^- \pi^0$ 
($\pi^0$ unobserved), responsible for a broad peak around $M_h \sim$ 500 MeV
(8.65\%);}
\item{\label{eta}
$q \to \eta\ X_5' \to \pi^+ \pi^- X_5$ with $X_5 = {\cal X}\ X_5'$: 
fragmentation into a $\eta(547)$ or $\eta'(958)$ decaying into 
$\pi^+ \pi^- {\cal X}$ (${\cal X}$ unobserved), responsible for a peak 
around $M_h \sim$ 350 MeV (2.05\%);}
\item{$q \to K^0\ X_6 \to \pi^+ \pi^- X_6$: fragmentation  into a $K^0$ resonance 
decaying into $\pi^+ \pi^-$, responsible for a narrow peak at $M_h \sim$ 498
MeV (3.41\%).}
\setcounter{saveenum}{\value{enumi}}
\end{enumerate}
On top of these, there could be the presence of two other channels:
\begin{enumerate}
\setcounter{enumi}{\value{saveenum}}
\item{$q \to \sigma \ X_7 \to \pi^+ \pi^- X_7$: 
fragmentation into the largely debated $\sigma$ resonance (see,
e.g., Ref.~\cite{Caprini:2005zr})  
decaying into $\pi^+ \pi^-$, which could be responsible for a very broad peak 
anywhere between 400 and 1200 MeV;}
\item{$q \to f_0 \ X_8 \to \pi^+ \pi^- X_8$: fragmentation into a $f_0$ resonance 
decaying into $\pi^+ \pi^-$, which should give rise to a peak at $M_h \sim$ 980 
MeV, not evident in the output of PYTHIA.}
\end{enumerate}
In our model, we considered only channels 1 to 6. 
All events not belonging to channels 2 to 6 were included in
channel 1, which then contains 70.77\% of the total events.

%%%%%%%%%%%%%%%%%%%%%%%%%%%%%%%%%
%\subsection{The spectator model}
%\label{s:correlator}

We work in the framework of a ``spectator'' model for
the fragmentation process: for $q\to \pi^+ \pi^- X$, the sum over
all possible intermediate states $X$ is replaced by an effective on-shell
state -- the spectator -- whose quantum numbers are in this case 
the same as the initial
quark and whose mass is one of the parameters of the model. In
principle, different channels could produce spectators with different masses. 
Moreover, each channel could end up into more than one possible
spectator~\cite{Kitagawa:2000ji}. 
For sake of simplicity, here we
consider just a single spectator for all channels. We shall 
denote its mass as $M_s$ and its momentum as $P_s$. 
The choice of using the same spectator for
all channels implies in particular that the fragmentation amplitudes of all
channels can interfere with each other maximally. 
In reality, it is plausible that only
a fraction of the total events ends up in the same spectator and can thus
produce interference effects.

Pions in channels 2 and 3 are obviously produced in relative $p$ wave, 
since they come
from the decay of a vector meson. 
In channel 4, each charged pion can be in a relative $p$ wave with respect to
the other one or to $\pi^0$, the net result being that there is a fraction of
$\pi^+ \pi^-$ pairs that is produced in a relative $s$ wave. In the following,
we will neglect this fraction and assume that all charged pairs are produced
in $p$ wave; at present we don't have
enough information to discriminate the two contributions. This assumption is
most probably inadequate and would lead to an overestimate of the contribution
of channel 4 to the final single spin asymmetry.
%and all pions in channel 4 are produced in $p$ wave.
%%This choice has the
%%consequence of giving rise to $sp$-wave interference in all the invariant-mass
%%range under consideration. 
%The latter assumption is most probably inadequate, because a fraction of the
%pions in channel 4 could be produced in $s$ wave. However, at present 
%we don't have
%enough information to separately study the two parts.
  
We further assume that all pions in channel 1 are produced in $s$ wave.
It is possible that a fraction of the background
events are also produced in $p$ wave. However, such a fraction cannot be too
big, as it would give rise to interference effects that would distort 
the shape of the $\rho$ meson peak. It is actually known that such a
distortion can indeed occur, but also that it is not
big~\cite{Lafferty:1993sx,Buskulic:1995gm}.  We think that this
point deserves further attention, but should not change the main features of our
results.

%We model the correlation function for $s$-wave and $p$-wave
%pairs in the following ways:
%\begin{align}
%\Delta^{q\to s}(k,P_h,R) &=
%\frac{1}{(2\pi)^4}\,
% \frac{(\kslash + m)}{k^2 - m^2} \, F^s\,e^{-\frac{k^2}{\Lambda_s^2}}\,
%(\kslash - \Pslash_h +M_s)\,
%%\\& \quad \times 
%F^{s \star}\,e^{-\frac{k^2}{\Lambda_s^2}}\,\frac{(\kslash + m)}{k^2 - m^2}\, 
% 2\pi\,\delta\bigl((k-P_h)^2 -M_s^2\bigr) \,,
%\\
%\Delta^{q\to p}(k,P_h,R) &=
%\frac{1}{(2\pi)^4}\,
% \frac{(\kslash + m)}{k^2 - m^2} \, F^p\,e^{-\frac{k^2}{\Lambda_p^2}}\,\Rslash\;
%(\kslash - \Pslash_h +M_s)\,\Rslash\;
%%\\& \quad \times 
%F^{p \star}\,e^{-\frac{k^2}{\Lambda_p^2}}\,\frac{(\kslash + m)}{k^2 - m^2}\, 
% 2\pi\,\delta\bigl((k-P_h)^2 -M_s^2\bigr) \,.
%\label{e:deltap}
%\end{align} 

\begin{figure}
\includegraphics[width=6cm]{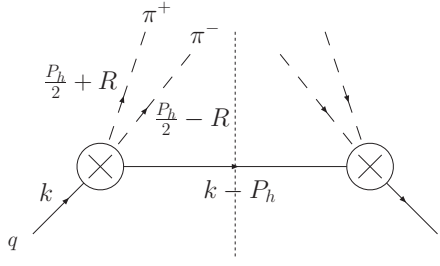}
\caption{
Diagrammatic representation of the correlation function $\Delta$ in the
spectator model.
}
\label{f:spect_fragm} 
\end{figure}

We model the correlation function in the following way (see Fig.~\ref{f:spect_fragm}):
\begin{equation} \begin{split} 
\Delta^{q}(k,P_h,R) &=
\frac{1}{(2\pi)^4}\,
 \frac{(\kslash + m)}{k^2 - m^2} \,\Bigl(F^{s \star}\,e^{-\frac{k^2}{\Lambda_s^2}} +
 F^{p \star}\,e^{-\frac{k^2}{\Lambda_p^2}}\,\Rslash\Bigr)
\,
(\kslash - \Pslash_h +M_s)\,
\\& \quad \times 
\Bigl(F^{s}\,e^{-\frac{k^2}{\Lambda_s^2}}+ F^{p}\,
e^{-\frac{k^2}{\Lambda_p^2}}\,\Rslash \Bigr)\,\frac{(\kslash + m)}{k^2 - m^2}\, 
 2\pi\,\delta\bigl((k-P_h)^2 -M_s^2\bigr) \,.
\label{e:deltamod}
\end{split} \end{equation}  
Isospin symmetry implies that 
the fragmentation correlator for $u \to \pi^+ \pi^- X$
is the same as for $\bar{d} \to \pi^+ \pi^- X$,  $d \to \pi^- \pi^+ X$,
$\bar{u} \to \pi^- \pi^+ X$. Therefore, the result for $d$ and $\bar{u}$ quarks
can be obtained from the result for $u$ quark by simply changing the sign
of $\vec{R}$, i.e. changing $\theta \to \pi - \theta$ and $\phi \to \phi +
\pi$. From now on we will drop the superscript indicating the quark flavor and
calculate the fragmentation functions for $u \to \pi^+ \pi^- X$.
The terms with vertex $F^{s}$ refer to the $s$-wave contribution, the terms
with vertex $F^{p}$ to the $p$-wave contribution.
The exponential form factors suppress the contributions from high quark
virtualities~\cite{Gamberg:2003eg}. 
%For simplicity, at some point 
%we will take the $\Lambda$ parameters to be the same in both 
%cases, but for the moment we keep them distinct. 
Other possibilities can be 
considered, e.g., dipole form factors~\cite{Bianconi:1999uc,Radici:2001na}, or sharp 
cutoffs~\cite{Bacchetta:2002tk}.

Inserting Eq.~(\ref{e:deltamod}) into Eq.~(\ref{eq:delta1}), we get  
\begin{equation} \begin{split}  
\Delta(z,\cos\theta,M_h^2,\phi_R) &=
\frac{|\vec R|}{128\,\pi^2\,M_h}\,\frac{z^2}{2\,(1-z)\,P_h^-}
\int d |\vec k_T|^2\,\biggl[
|F^s|^2\, e^{-\frac{2\,k^2}{\Lambda_s^2}}\,\frac{(\kslash + m)\,
(\kslash - \Pslash_h +M_s)\,(\kslash + m)}{(k^2 - m^2)^2}
\\ &\quad +
|F^p|^2\, e^{-\frac{2\,k^2}{\Lambda_p^2}}\,\frac{(\kslash + m)\,\Rslash\;
(\kslash - \Pslash_h +M_s)\,\Rslash\;\,(\kslash + m)}{(k^2 - m^2)^2}
\\ &\quad +
F^{s \star}\,F^{p}\, e^{-k^2\,\frac{\Lambda_s^2+\Lambda_p^2}{\Lambda_s^2\,\Lambda_p^2}}
\,\frac{(\kslash + m)\,(\kslash - \Pslash_h +M_s)\,\Rslash\;\,(\kslash + m)}
{(k^2 - m^2)^2}
\\ 
&\quad + F^{s}\,F^{p \star}\, 
e^{-k^2\,\frac{\Lambda_s^2+\Lambda_p^2}{\Lambda_s^2\,\Lambda_p^2}}\,
\frac{(\kslash + m)\,\Rslash\; (\kslash - \Pslash_h +M_s)\,(\kslash + m)}
{(k^2 - m^2)^2} \biggr], \label{e:deltamod2}
\end{split} \end{equation}  
with $k^2$ fixed by the on-shell condition of the
spectator~\cite{Bianconi:1999uc}, i.e., 
\begin{equation}
k^2 = \frac{z}{1-z}\,|\vec k_T|^2+\frac{M_s^2}{(1-z)}+\frac{M_h^2}{z} .
\end{equation}
The first and second lines of Eq.~(\ref{e:deltamod2}) describe the pure $s$- and 
$p$-wave contributions
and, as such, they are the only ones that can contribute to the functions
$D_{1,oo}$, $D_{1,ll}$ of Eq.~(\ref{eq:d1pw}) and $H_{1,lt}^{\open}$ of
Eq.~(\ref{eq:h1angpw}), while the third and fourth
lines describe the $sp$ interference and they contribute to the functions
$D_{1,ol}$ and $H_{1,ot}^{\open}$.

For convenience, we introduce the function
\begin{equation}
L^2(z,M_h^2) = \frac{1-z}{z^2}\,M_h^2 +\frac{1}{z}\,M_s^2 -
\frac{1-z}{z}\,m^2.  
\end{equation} 
This function has to be always positive for kinematical reasons.
%, leading in
%particular to the condition $M_h-m\leq M_s \leq M_h+m$.

We obtain the following result for the unpolarized fragmentation function
\begin{equation} \begin{split}
D_{1,oo}(z,M_h^2) &=  \frac{z\,|\vec R|}{16\, \pi\,M_h}\,|F^s|^2\, e^{-\frac{2\,m^2}{\Lambda_s^2}}\,
\Biggl[\biggl(1+2\,\frac{M_h^2-(m+M_s)^2}{z\,\Lambda_s^2}\biggr)
        \,\Gamma\biggl(0,\frac{2\,z\,L^2}{(1-z)\,\Lambda_s^2}\biggr) 
\\ & \quad 
- \frac{1-z}{z^2}\,\frac{M_h^2-(m+M_s)^2}{L^2}\, e^{-\frac{2\,z\,L^2}{(1-z)\,\Lambda_s^2}}\Biggr] 
\\ & \quad 
+ \frac{z\,|\vec R|}{16\, \pi\,M_h}\,|F^p|^2\, e^{-\frac{2\,m^2}{\Lambda_p^2}}\,
\frac{|\vec R|^2}{3\,M_h^2}
%\\& \times 
\,\Biggl[\biggl(2\,M_h^2+\frac{2-z}{z}\,(m^2-M_s^2)
\\ & \quad 
+ 2\,\frac{(M_h^2 - (m-M_s)^2)\,(2\,M_h^2 + (m+M_s)^2)}{z\,\Lambda_p^2}
\biggr)\,\Gamma\biggl(0,\frac{2\,z\,L^2}{(1-z)\,\Lambda_p^2}\biggr)
\\ & \quad  
+\frac{1-z}{2\,z^2\,L^2}\,\biggl(\Bigl(M_s^2 + \frac{1-z}{z}\,(M_h^2-z\, m^2)\Bigr)\Lambda_p^2
\\ & \quad
-2\,(M_h^2 - (m-M_s)^2)\,(2\,M_h^2 + (m+M_s)^2)\biggr)\, e^{-\frac{2\,z\,L^2}{(1-z)\,\Lambda_p^2}}
\Biggr]. 
\end{split} \end{equation}  
The incomplete $\Gamma$ function -- typically appearing in model calculations
with exponential form-factors~\cite{Gamberg:2003eg} --  
is defined as $\Gamma(0,z) \equiv
\int_z^\infty e^{-t}/t\, dt$. 
%If the fragmenting quark is taken to be
%massless (implying also $M_s=M_h$), the above expression simplifies
%considerably and becomes
%\begin{equation} \begin{split}
%D_{1,oo}(z,M_h^2) &\approx \frac{z\,|\vec R|}{16\,\pi\,M_h}\,\Biggl\{
%|F^s|^2 \,\Gamma\biggl(0,\frac{2\,M_h^2}{z\,(1-z)\,\Lambda_s^2}\biggr) 
%\\ & \quad 
%+ |F^p|^2\,|\vec R|^2\,
%\,\Biggl[\Bigl(1-\frac{2}{3\,z}\Bigr)
%\,\Gamma\biggl(0,\frac{2\,M_h^2}{z\,(1-z)\,\Lambda_p^2}\biggr)
%%\\ & \quad  
%+ \frac{1-z}{6\,z}\,
%\frac{\Lambda_p^2}{M_h^2}\,
%e^{-\frac{2\,M_h^2}{z\,(1-z)\,\Lambda_p^2}}\Biggr]
%\Biggr\}. 
%\end{split} \end{equation}  
The first term of the fragmentation function can be identified with the pure
$s$-wave contribution, also called $D_{1,oo}^{s}/4$ in
Ref.~\cite{Bacchetta:2002ux}, and the second one with the pure $p$-wave
contribution, also called $3\, D_{1,oo}^{p}/4$.

We give here the results also for the other functions appearing in
Eq.~(\ref{eq:d1pw}):
\begin{align}
\begin{split}
D_{1,ol}(z,M_h^2) &= \frac{z |\vec R|}{16 \pi\,M_h}\,2\,{\rm Re}(F^{s \star} F^p)\, 
e^{-\frac{2\,m^2}{\Lambda_{sp}^2}}
%e^{-m^2\frac{\Lambda_s^2+\Lambda_p^2}{\Lambda_s^2\Lambda_p^2}}
\frac{|\vec R|}{z^2 M_h}\,
\Biggl[(m+M_s)\,\frac{1-z}{z}\,\frac{(2-z)\,M_h^2 - z\,(m^2-M_s^2)}{L^2}\,
e^{-\frac{2\,z\,L^2}{(1-z)\,\Lambda_{sp}^2}}
%e^{-\frac{z L^2}{(1-z)}\,\frac{\Lambda_s^2+\Lambda_p^2}{\Lambda_s^2 \Lambda_p^2}}
\\ & \quad 
-\biggl(2\,(m+M_s)\,\frac{(2-z)\,M_h^2 -
  z\,(m^2-M_s^2)}{\Lambda_{sp}^2}+z\bigl(m+(1-z)M_s\bigr)\biggr)\Gamma\biggl(0,\frac{2\,z\,L^2}{(1-z)\,\Lambda_{sp}^2}\biggr)
%\Gamma\biggl(0,\frac{z\,L^2}{(1-z)}\,\frac{\Lambda_s^2+\Lambda_p^2}{\Lambda_s^2\,\Lambda_p^2}\biggr)
\Biggr],
\label{e:d1ol}
\end{split}
\\
\begin{split}
D_{1,ll}(z,M_h^2) &=
%\frac{z\,|\vec R|}{16\, \pi\,M_h}\,|F^p|^2\, e^{-\frac{2\,m^2}{\Lambda_p^2}}\,
%\frac{3\,|\vec R|^2}{4\,M_h^2}
%\,\Biggl\{
%\biggl[2\,M_h^2+\frac{2-z}{z}\,(m^2-M_s^2)
%%\\ & \quad 
%+ 2\,\frac{(M_h^2 - (m-M_s)^2)\,(2\,M_h^2 + (m+M_s)^2)}{z\,\Lambda_p^2}
%\\ & \quad 
%-\frac{3\,M_h^2}{z^3\,\Lambda^2}\,\Bigl(\bigl(2-2z+z^2\bigr)\bigl(z\Lambda^2+2
%M_h^2\bigr)+
%4 z^2\,
%m\, M_s-2z\,(2-z)\bigl(m^2-M_s^2\bigr)\Bigr)\biggr]
%\,\Gamma\biggl(0,\frac{2\,z\,L^2}{(1-z)\,\Lambda_p^2}\biggr)
%\\ & \quad  
%+\frac{1-z}{2\,z^2\,L^2}\,\biggl[\Bigl(M_s^2 + \frac{1-z}{z}\,(M_h^2-z\, m^2)\Bigr)\Lambda_p^2
%%\\ & \quad
%-2\,(M_h^2 - (m-M_s)^2)\,(2\,M_h^2 + (m+M_s)^2)\biggr)\, 
%\\ & \quad
%+\frac{3\,M_h^2}{z^2}\,\Bigl(\bigl(2-2z+z^2\bigr)\bigl(z\Lambda^2+2
%M_h^2\bigr)+
%4 z^2\,
%m\, M_s-2z\,(2-z)\bigl(m^2-M_s^2\bigr)\Bigr)\biggr]
%e^{-\frac{2\,z\,L^2}{(1-z)\,\Lambda_p^2}}
%\Biggr\}.
%\\
%&=
\frac{27}{16}\,D_{1,oo}^{p}(z,M_h^2) + \frac{z\,|\vec R|}{16\, \pi\,M_h}\,|F^p|^2\, e^{-\frac{2\,m^2}{\Lambda_p^2}}\,
\frac{9\,|\vec R|^2}{4\,z^3\,\Lambda_p^2}\,\biggl\{
\biggl[\frac{1-z}{2 z L^2}\,\Lambda_p^2\,e^{-\frac{2\,z\,L^2}{(1-z)\,\Lambda_p^2}}
-\Gamma\biggl(0,\frac{2\,z\,L^2}{(1-z)\,\Lambda_p^2}\biggr)\biggr]
\\ & \quad \times
\Bigl(2 M_h^2 \bigl(2-2z+z^2\bigr)+
4 z^2\,
m\, M_s-2z\,(2-z)\bigl(m^2-M_s^2\bigr)\Bigr)
- z\,\Lambda_p^2\,
\bigl(2-2z+z^2\bigr)
\Gamma\biggl(0,\frac{2\,z\,L^2}{(1-z)\,\Lambda_p^2}\biggr)
\biggr\},
\end{split}
\end{align} 
where $2/\Lambda_{sp}^2 = 1/\Lambda_{s}^2 + 1/\Lambda_{p}^2$.
%\begin{equation}
%\frac{2}{\Lambda_{sp}^2} =  \frac{1}{\Lambda_{s}^2} + \frac{1}{\Lambda_{p}^2}.
%\end{equation} 

For the interference fragmentation function $H_{1,ot}^{\open}$ we obtain
\begin{equation} \begin{split} 
H_{1,ot}^{\open}(z,M_h^2) &= -\frac{z\,|\vec R|}{16\, \pi\,M_h}\,2\,{\rm
  Im}(F^{s \star} \, F^p)\, 
e^{-\frac{2\,m^2}{\Lambda_{sp}^2}}\,
%e^{-m^2\,\frac{\Lambda_s^2+\Lambda_p^2}{\Lambda_s^2\,\Lambda_p^2}}
\,\frac{M_h}{z^2}\,
%\\ & \times 
\Biggl[
\frac{1-z}{z}\,\frac{M_h^2-z^2\,m^2}{L^2}\,
e^{-\frac{2\,z\,L^2}{(1-z)\,\Lambda_{sp}^2}}
%e^{-\frac{z\,L^2}{(1-z)}\,\frac{\Lambda_s^2+\Lambda_p^2}{\Lambda_s^2\,\Lambda_p^2}}
\\ & \quad 
-\biggl(z +2\,\frac{M_h^2-z^2\,m^2}{\Lambda_{sp}^2}\biggr)\,
%-\biggl(z+(M_h^2-z^2\,m^2)\,\frac{\Lambda_s^2+\Lambda_p^2}{\Lambda_s^2\,\Lambda_p^2}\biggr)\,
        \Gamma\biggl(0,\frac{2\,z\,L^2}{(1-z)\,\Lambda_{sp}^2}\biggr)
%\Gamma\biggl(0,\frac{z\,L^2}{(1-z)}\,\frac{\Lambda_s^2+\Lambda_p^2}{\Lambda_s^2\,\Lambda_p^2}\biggr) 
\Biggr].
%\\ &\stackrel{m\to 0}{\approx}
%\frac{z\,|\vec R|}{16\,\pi\,M_h}
%\,2\,{\rm Im}(F^{s \star} \, F^p)\,\frac{M_h}{z^2}
%\Biggl[\biggl(z+M_h^2\,\frac{\Lambda_s^2+\Lambda_p^2}{\Lambda_s^2\,\Lambda_p^2}\biggr)\,
%        \Gamma\biggl(0,\frac{M_h^2}{z\,(1-z)}\,\frac{\Lambda_s^2+\Lambda_p^2}{\Lambda_s^2\,\Lambda_p^2}\biggr) 
%\\ & \quad 
%-
%z\,(1-z)\,e^{-\frac{M_h^2}{z\,(1-z)}\,\frac{\Lambda_s^2+\Lambda_p^2}{\Lambda_s^2\,\Lambda_p^2}}
%\Biggr] 
\label{e:h1angle}
\end{split} \end{equation}  
%%%%%%%%%%%%%%%%%%%%%%%%%%%%%%%%%%%%%%%%%%%%%%%%%%%%

The interference function $H_{1,lt}^{\open}$ vanishes in our model, since 
it would be
proportional to ${\rm
  Im}(F^{p \star} \, F^p)$.
It would be
necessary to have other sources of nontrivial phases 
in the amplitudes for $p$ wave production, as the ones given by loop
corrections~\cite{Amrath:2005gv}.

The vertices $F^s$ and $F^p$ are essential ingredients to
reproduce the correct invariant-mass behavior and to produce the imaginary
parts required for the generation of T-odd fragmentation functions. When no
resonance is present, we assume the vertices to be real. When resonances
are present, the vertex includes the resonance propagator
and contains therefore real and imaginary parts.  
We assume in our calculation that the $s$-wave background is free of
resonances and thus is purely real (at tree level). 
This is one of the main assumptions we
make in the present work and has critical consequences on the invariant-mass
behavior of the fragmentation functions $D_{1,ol}$ and  $H_{1,ot}^{\open}$.  
As can be readily seen from Eqs.~(\ref{e:d1ol}) and (\ref{e:h1angle}), 
assuming $F^s$ to be real 
implies that $D_{1,ol}$ is proportional to ${\rm Re}(F^p)$  and
$H_{1,ot}^{\open}$ is proportional to ${\rm Im}(F^p)$.

At this point it is worthwhile
to make a comparison with the results presented in the literature in the past.
In
Ref.~\cite{Jaffe:1998hf}, the necessary phase difference between $s$ and
$p$ waves was taken from $\pi\,\pi$ phase shifts in elastic scattering 
data~\cite{Estabrooks:1974vu}.  
No hypothesis was made
on the invariant-mass behavior of the $s$
and $p$ amplitudes, i.e., on the fragmentation mechanism. 
The main assumption was that the
interference pattern occurring in the fragmentation process, where the initial state is a quark and the final state is composed of
many hadrons, is supposed to be the same as in $\pi\,\pi$ scattering, where initial and
final states are simply two pions.
In particular, the prediction of Ref.~\cite{Jaffe:1998hf} 
changes sign close to the $\rho$ mass. The effect is in fact proportional to
the sum of 
the real part of the $\rho$ resonance times the imaginary part of the $\sigma$
plus the real part of the $\sigma$ times the imaginary part of the
$\rho$. Both real parts change sign close to the $\rho$ mass. 
%This real part has to
%interfere with an imaginary component in the $s$ wave (e.g.\
%coming from the $\sigma$, channel 7 in our list).
%Thus, in Ref.~\cite{Jaffe:1998hf}
%the invariant-mass behavior of the T-odd
%fragmentation function $H_{1,ot}^{\open}$ 
%received contributions from the interference between
%the imaginary part of the $\rho$ resonance (which doesn't change sign at the
%$\rho$ mass peak) times the real part of the $\sigma$ (which changes sign at the
%$\sigma$ mass), as well as from the interference between
%the real part of the $\rho$ resonance (which changes sign at the
%$\rho$ mass peak) times the imaginary part of the $\sigma$ (which doesn't change 
%sign at the
%$\sigma$ mass). 
%Phase-shift studies give direct information on the combination of these two
%contributions. In particular, they display a sign change close
%to the $\rho$ mass peak~\cite{Jaffe:1998hf}. 
On the contrary, in our approach we neglect the contribution of the $\sigma$
and we take a purely real $s$-wave background, but 
we try to take into account in a collective manner 
all other ways in which pion pairs can be
produced in a semi-inclusive fragmentation process.

A different model prediction was also presented in
Ref~\cite{Radici:2001na}. In that model, the $s$-wave amplitude was modeled as
a sequence of two single-pion emissions and was purely real, while the
$p$-wave amplitude contained only the $\rho$ decay, therefore having a
predictive power limited to invariant masses around the $\rho$ resonance. 
The $s-p$ phase difference 
followed from
the presence of an imaginary part in the meson propagator. The parameters of
the model were fixed using theoretical arguments, since no experimental input
was available as a comparison. 
As already pointed out, 
the present work is similar to Ref~\cite{Radici:2001na}, but
the $s$-wave amplitude is modeled by means of an effective real vertex
and the $p$-wave amplitude contains contributions from the $\rho$ and the
$\omega$ mesons. Last but not least, 
the parameters of the model are fixed by fitting
the output of the PYTHIA Monte Carlo generator, which is known to reproduce
very well the unpolarized data.

Our ansatz for the vertices is
\begin{align}
F^s &= f_{s},
\\
\begin{split} 
F^p &= f_{\rho}\,\frac{(M_h^2 - M_{\rho}^2)- i\, \Gamma_{\rho}\,M_{\rho}}
        {(M_h^2 - M_{\rho}^2)^2 + \Gamma_{\rho}^2\,M_{\rho}^2}
%\\ & \quad 
+ f_{\omega}\,\frac{(M_h^2 - M_{\omega}^2)- i\, \Gamma_{\omega}\,M_{\omega}}
        {(M_h^2 - M_{\omega}^2)^2 + \Gamma_{\omega}^2\, M_{\omega}^2}
\\ & \quad
%be careful with the definition of this sign!
-
i\,f'_{\omega}\,\frac{\sqrt{\lambda\bigl(M_{\omega}^2,M_h^2,m_{\pi}^2\bigr)}\,
\Theta(M_{\omega}-m_{\pi}-M_h)}
        {4\,\pi\,\Gamma_{\omega}\,M_{\omega}^2\,
        \sqrt[4]{4\,M_{\omega}^2\,m_{\pi}^2
        +\lambda\bigl(M_{\omega}^2,M_h^2,m_{\pi}^2\bigr)}},
\label{e:Fp}
\end{split}
\end{align} 
where $\lambda\bigl(M_{\omega}^2,M_h^2,m_{\pi}^2\bigr) = [M_{\omega}^2 -
(M_h+m_{\pi})^2][M_{\omega}^2 - (M_h-m_{\pi})^2]$ and $\Theta$ denotes the
unit step function.
The couplings $f_{s}$, $f_{\rho}$, $f_{\omega}$ and $f'_{\omega}$
are parameters of the model. The first two terms of $F^p$ can be easily
identified with the contributions of the $\rho$ and the $\omega$ resonances
decaying into two pions. The Lorentz structure of the resonance 
propagators is already taken into account in Eq.~(\ref{e:deltamod}). The masses
and widths of the two resonances are taken from the PDG~\cite{Eidelman:2004wy}: $M_{\rho} = 0.776$
GeV, $\Gamma_{\rho} = 0.150$ GeV, $M_{\omega} = 0.783$
GeV, $\Gamma_{\omega}  = 0.008$ GeV. The details of the
resonance propagators could be also extracted from phase-shift analyses, as
done in Ref.~\cite{Jaffe:1998hf}. In this case the contribution of the $\rho$
resonance would be
\begin{equation} 
f_{\rho}\,\frac{-\tan{\delta_1^1} - \ii\,
  \tan^2{\delta_1^1}}{\Gamma_{\rho}\,M_{\rho}\,(1+\tan^2{\delta_1^1})},
\end{equation} 
where $\delta_m^l$ are the $\pi \pi$ phase shifts for the specific $(l,m)$ channel. 
However, using $\delta_m^l$ from, e.g., Ref.~\cite{Colangelo:2001df} leads to no
significant change compared to Eq.~(\ref{e:Fp}), especially considering 
the coarse level of accuracy of our model.

The third
term in Eq.~(\ref{e:Fp}) comes from the decay of an $\omega$ 
resonance into three pions, of which the
$\pi^0$ goes undetected, i.e.,  channel \ref{omega3pi}. 
Let's introduce the momentum $P_{\omega} = P_h + P_0$, where $P_0$ is the
momentum of $\pi^0$ and $P_{\omega}^2 = M_3^2$.
Taking for $\omega$ the usual Lorentz structure of a vector
meson resonance, the evaluation of the correlator $\Delta$ involves the
contraction
\begin{equation}
\gamma_{\mu} \biggl(-g^{\mu \nu} + \frac{P_{\omega}^{\mu} P_{\omega}^{\nu}}{M_3^2}
\biggr)R_{\nu} = -\Rslash + \frac{R\cdot P_0\, (\Pslash_h
  +\Pslash_0)}{M_3^2} =
-\Rslash\,\biggl(1+\frac{|\vec{P}_0|}{3 M_3^2}\biggr).
\label{e:boh}
\end{equation} 
In the last step we took into account the fact that $R$ has no timelike
component and we have to integrate over $\vec{P}_0$. 
When performing the integration over $P_0$, we make a narrow-width
approximation  and we assume that the three pions are produced exactly at the
$\omega$ mass ($M_3 = M_{\omega}$). In this approximation, $|\vec{P}_0|$ is fixed,
the last term of Eq.~(\ref{e:boh}) becomes negligible,
% I
%After integrating over 
%The above invariant-mass behavior is obtained by assuming that 
%the three pions are produced exactly at the $\omega$ mass (narrow-width approximation). The $\pi^0$ variables  
%are then integrated. 
the contribution of channel \ref{omega3pi} to the vertex 
turns out to be purely
imaginary and occurs only at invariant masses lower than
$M_{\omega}-m_{\pi} = 0.643$ GeV. Abandoning 
the narrow-width approximation has the consequence 
of smearing the invariant-mass distribution and allowing pairs to appear at
invariant masses higher than 0.643 GeV, as well as giving a real part to the
third term of Eq.~(\ref{e:Fp}).

Note that we sum the three contributions in Eq.~(\ref{e:Fp}) at the amplitude
level. This is the first instance where the assumption of equal spectators has
a direct consequence, and deserves further comments.
Channels \ref{rho}, \ref{omega2pi} and
\ref{omega3pi} can interfere if $X_2=X_3=X_4$. In general, only a
fraction of the total events will fulfill this requirement.
On the contrary, since we have a single spectator for all channels 
this is {\em always} the case in our model. That's why we add up the
amplitudes in Eq.~(\ref{e:Fp}).
As we shall see in the next section, 
the effect of these interferences in the unpolarized fragmentation function 
is in any
case quite small, 
due to the small contribution of channel \ref{omega2pi} and the
small overlap between channels \ref{rho} and \ref{omega3pi}. However, a
similar problem will show up also in the calculation of the function
$H_1^{\open}$, with more relevant consequences, 
as we shall see in Sec.~\ref{s:results}.  

Finally, we felt the need to use $z$-dependent $\Lambda$-cutoffs to have an
acceptable description of the data. We used the following ansatz:
\begin{equation}
\Lambda_{s,p}= \alpha_{s,p}\, z^{\beta_{s,p}}\,(1-z)^{\gamma_{s,p}}. 
\end{equation} 

The total number of parameters of the model is thus 12: 
3 parameters for the form-factor 
cutoff $\Lambda_s$, 3 parameters for the
cutoff $\Lambda_p$, the couplings
$f_{s}$, $f_{\rho}$, $f_{\omega}$ and $f'_{\omega}$, the mass of the
spectator, $M_s$,
and the mass of the
fragmenting quark, $m$. However, in the following we shall always assume $m=0$.
 
%%%%%%%%%%%%%%%%%%%%%%%%%%%%%%%%%%%%%%%%%%%%%%%%%%%%%%%%%%%%%%%%%%%%%%%%%%%%%
\section{Comparison with PYTHIA and parameters fit}
\label{s:d1}

In order to fix the parameters of the model, we compare it to the output of
the PYTHIA event generator~\cite{Sjostrand:2000wi} tuned for HERMES~\cite{Liebing:2004us}.
The events are generated in
4$\pi$. 
Exclusive channels are dropped. 
The standard HERMES 
semi-inclusive DIS cuts
are applied, in particular $Q^2>1$ GeV$^2$, $0.1 <y< 0.85$, $0.023 <x< 0.4$, 
$W^2> 4$ GeV$^2$
and the momenta of the pions ($|\vec{P}_{1,2}|$) are 
constrained to be larger than 1 GeV.
\footnote{To perform the fit, we neglected the last cut.} 
The counts
per $M_h$-bin 
are proportional to the cross section of Eq.~(\ref{eq:crossOO}) times $2
M_h$ (since the cross section in the former equation is differential 
in $d M_h^2$),  
integrated over $y$, $x$, $\phi_R$, $\phi_S$, and further over
$0.2 < z < 0.8$. For the counts per $z$-bin, we integrated the cross
section over $0.3\,\text{GeV} < M_h < 1.3\,\text{GeV}$.

\begin{figure}
\begin{tabular}{ccc}
\includegraphics[width=8cm]{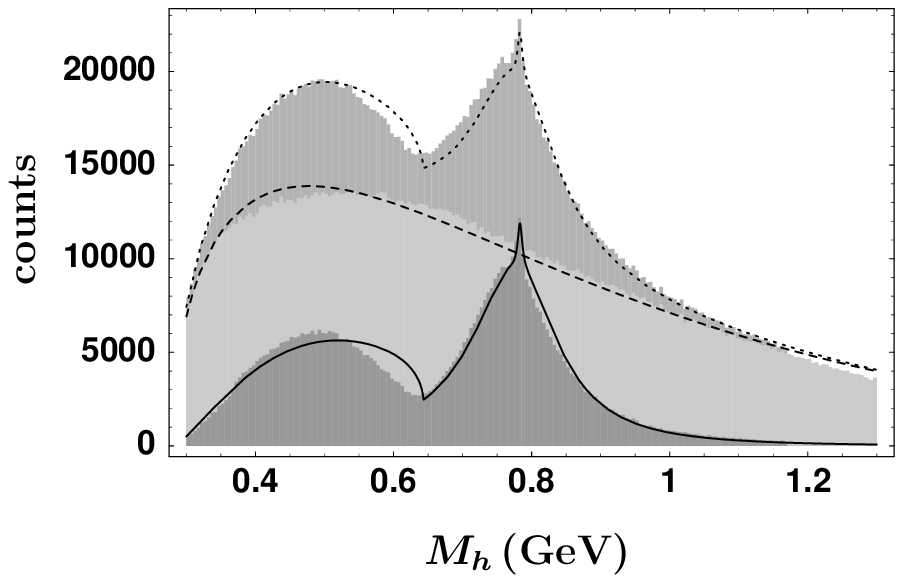}
&\hspace{1cm} &
\includegraphics[width=8cm]{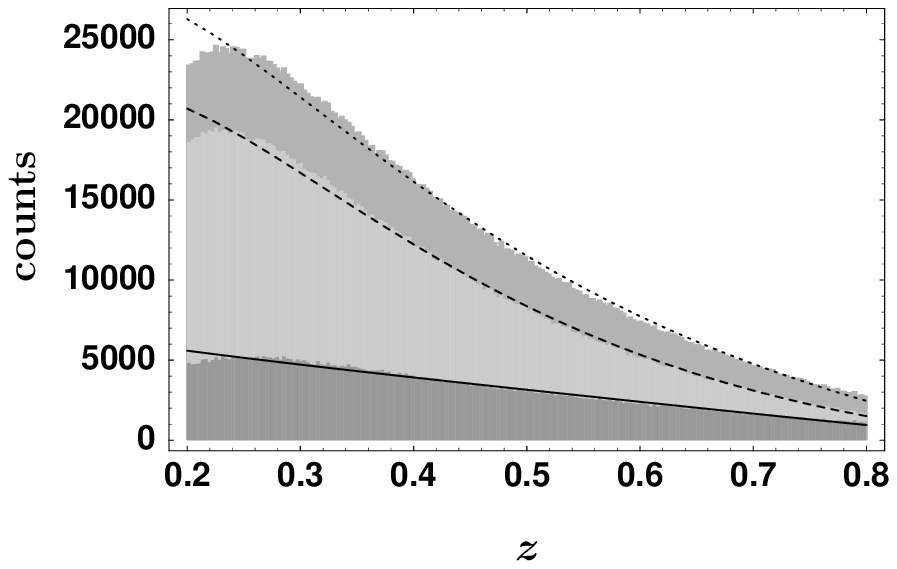}
\\
(a) & & (b)
\end{tabular}
\caption{Semi-inclusive dihadron counts from the PYTHIA event
  generator~\cite{Sjostrand:2000wi}
 tuned
  for HERMES~\cite{Liebing:2004us}
 and results of the fit (a) as a function of $M_h$, (b)
as a function of $z$. Solid line: $p$-wave contribution; dashed line:
$s$-wave contribution; dotted line: sum of the two. The contributions of the
$\eta$ and $K^0$ have been excluded.}
\label{f:fit}
\end{figure}

In Fig.~\ref{f:fit} the number of counted dihadron pairs is presented
binned in $M_h$ (200 bins from 0.3 to 1.3 GeV) and $z$ (200 bins from 0.2 to 0.8).
From the total counts, we excluded the contributions
from $\eta$ and $K^0$, i.e., channels 5 and 6 (see Fig.~\ref{f:pythia}),
because they are not relevant for our purposes.
The lowest dark-gray histogram represents the 
sum of the $\rho$ and $\omega$ contributions (channels 2, 3, and 4), assumed
to describe the whole $p$-wave contribution. The light-gray histogram in the
middle 
is the ``background'' contribution,
i.e. channel 1, representing the  
$s$-wave contribution.
The upper histogram is the sum of the other two and corresponds to the total
counts minus channels 5 and 6.

%To fit the parameters of the model, we considered separately the background
%(channel 1) -- described by the $s$-wave part of our model -- 
%and the sum of channels 2, 3, and 4 -- described by the $p$-wave part of the
%model. 
Instead of leaving all parameters of the model free, 
for sake of simplicity we assumed the
fragmenting quark to be massless. 
%which in its turn implied that the mass of
%the spectator is also fixed by the kinematics to be equal to $M_h$. 
We take the spectator mass to be proportional to the pair's invariant mass.
The number of free parameters we
used is then 11.
%, of which 4 for the $s$-wave channel, 6 for the $p$-wave
%channel

The $\chi^2$ minimization was performed using 
 MINUIT. The $\chi^2$ function was
defined as the square of the difference between the expected number of events
in the bin and the
measured value, divided by the expected number (equivalent to assigning a
statistical error equal to the square root of the number of events in the
bin). The resulting $\chi^2/\text{d.o.f}$ is very high, about 25. 
However, we
believe that the main characteristics of the $M_h$ and $z$ shapes of the
unpolarized fragmentation functions are qualitatively well described.
The result of the fit for the $s$ and $p$ wave 
is shown on top of the PYTHIA output in Fig.~\ref{f:fit}.

The values of the parameters obtained by the fit are:
\begin{align}
\alpha_s &=2.60 \pm 0.05\; \text{GeV}^{2},
&
\beta_s &=-0.751 \pm 0.008,
&
\gamma_s &=-0.193 \pm 0.004,
\\
\alpha_p &=7.07 \pm 0.11\; \text{GeV}^{2},
&
\beta_p &=-0.038 \pm 0.003,
&
\gamma_p &=-0.085 \pm 0.004,
\\
f_{s}&=1197 \pm 2 \; \text{GeV}^{-1},
&
f_{\rho}&=93.5 \pm 1.6,
&
f_{\omega}&=0.63 \pm 0.03,
\\
f'_{\omega}&=75.2 \pm 1.2,
&
M_s &= (2.97 \pm 0.04)\,M_h.
\end{align} 
The coupling constants are fixed modulo an overall 
normalization factor which depends on the luminosity and 
is irrelevant for asymmetry calculations. The sign of the coupling constants
is also {\em not} fixed, but the relative sign of  $f_{\rho}$,
$f_{\omega}$ and $f'_{\omega}$  is (see below).

In the $p$-wave channel, our model deviates significantly from the generated
spectrum in the region around 0.6 GeV, substantially increasing the
$\chi^2$. This is due to the interference between channels 2 and
4, which is not included in the Monte Carlo generator. 
At the same time, in the $s$-channel the curve
obtained from our model underestimates the data in the same region. 
Thus, the sum of the two curves is in good agreement with the total
generated spectrum, to which the Monte Carlo generator is actually
tuned. The agreement would be improved further if the
contribution of the $\omega$ were extended at higher invariant masses by
leaving the narrow-width approximation for the $\omega$ resonance
and smearing the step function in Eq.~(\ref{e:Fp}).
% 
%On the other hand, the disagreement with the Monte Carlo output in the
%$p$-channel could be
%an indication that
%the interference between the $\rho$ and $\omega$ 
%channels is much smaller in reality than in
%our model. 
%As mentioned before, this could be taken into account assuming that
%only a certain fraction of the total events interferes. However, we refrain
%ourselves from adding another parameter to the model.
Note that the interference is in this case
constructive because the signs of the couplings $f_{\rho}$ and $f'_{\omega}$
have been taken equal. 
If the two couplings were taken opposite, then a
destructive interference would take 
place and the model would underestimate
the $p$-wave data at around 0.6 GeV. The agreement with the total spectrum
would then be worsened. Also the $f_{\omega}$ coupling has been taken to
have the same sign of $f_{\rho}$ to avoid destructive interference patterns.
It is difficult with the present poor knowledge 
to make any conclusive statement about $\rho$-$\omega$
interference in semi-inclusive dihadron production.
However, we can at least 
conclude that in our model the best agreement with the event
generator is achieved when the three couplings 
$f_{\rho}$, $f_{\omega}$ and $f'_{\omega}$ have the same sign.

%%%%%%%%%%%%%%%%%%%%%%%%%%%%%%%%%%%%%%%%%%%%%%%%%%%%%%%%%%%%%%%%%%%%%%%%%%%%%
\section{Predictions for polarized fragmentation functions and 
transverse-spin asymmetry}
\label{s:results}

\begin{figure}
\begin{tabular}{ccc}
\includegraphics[width=7cm]{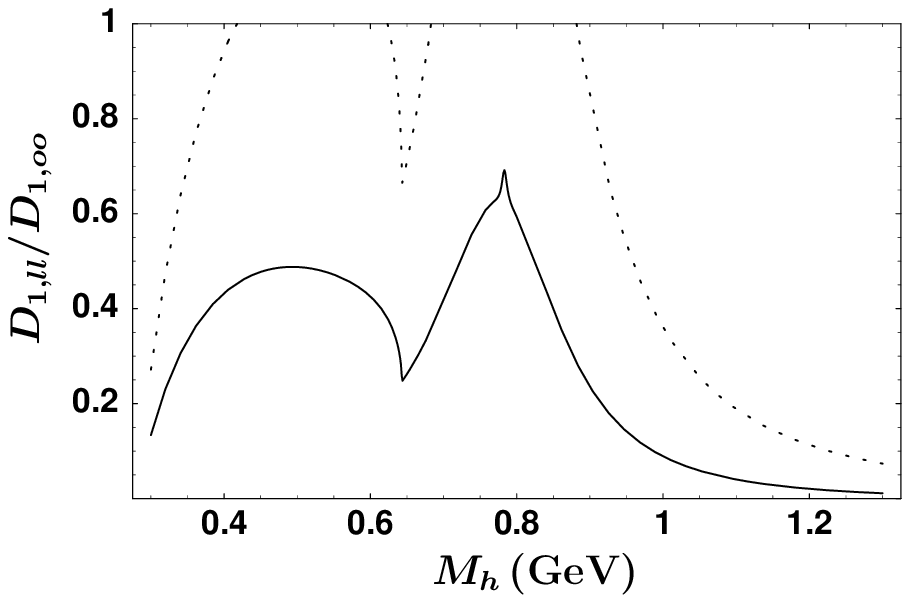}
&\hspace{1cm} &
\includegraphics[width=7cm]{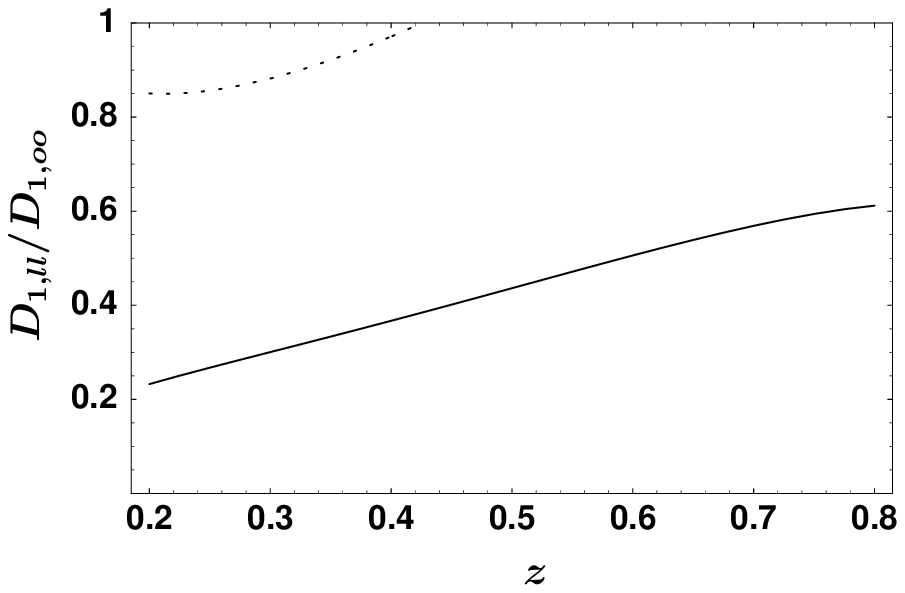}
\\
(a) & & (b)
\end{tabular}
\caption{Model prediction for the ratio $D_{1,ll}/
  D_{1,oo}$: (a) as a function of $M_h$, (b) as a function of $z$. 
  The dotted lines represent the positivity bounds of Eq.~(\ref{e:posd1ll}).} 
\label{f:d1ll}
\end{figure}

Using the parameters obtained from the fit we can plot the results for the
fragmentation 
functions  
$D_{1,ll}$, 
%Eq.~(\ref{e:d1ll}),
$H_{1,ot}^{\open}$,
%, Eq.~(\ref{e:h1angle}),
and
$D_{1,ol}$.
% Eq.~(\ref{e:d1ol}).
The function $D_{1,ll}$ is a pure $p$-wave function. It depends on $|F^p|^2$,
the modulus square of Eq.~(\ref{e:Fp}), and has a behavior very similar to
$D_{1,oo}^{p}$, the $p$-wave part of  $D_{1,oo}$. 
In Fig.~\ref{f:d1ll} (a) we plot the ratio between 
$D_{1,ll}$
and $D_{1,oo}$, integrated separately 
over $0.2< z < 0.8$. In Fig.~\ref{f:d1ll} (b) we plot the same ratio
but with the two functions multiplied by $2 M_h$ and integrated over $0.3\,\text{GeV} < M_h < 1.3\,\text{GeV}$. In the same
figures, the dotted lines represent 
the positivity
bound~\cite{Bacchetta:2002ux}
\begin{equation}
-\frac{3}{2}\,D_{1,oo}^p \leq D_{1,ll} \leq 3 D_{1,oo}^p.
\label{e:posd1ll}
\end{equation} 

The functions $D_{1,ol}$ and $H_{1,ot}^{\open}$ arise 
from the interference of $s$ and $p$ waves,
i.e.\ from the interferences of channels 1-2, 1-3, and  1-4, 
proportional to the 
product $(f_{s}\, f_{\rho})$, $(f_{s}\, f_{\omega})$, $(f_{s}\, f'_{\omega})$,
respectively.  
Since the relative sign of $f_{s}$ and the $p$-wave couplings is
not fixed by the fit, we can only predict
these functions modulo a sign. For the plots, we assume that the $p$-wave
couplings have a sign {\em opposite} to $f_{s}$ (as suggested by the sign of
preliminary HERMES data~\cite{vanderNat:2005yf}).

\begin{figure}
\begin{tabular}{ccc}
\includegraphics[width=7cm]{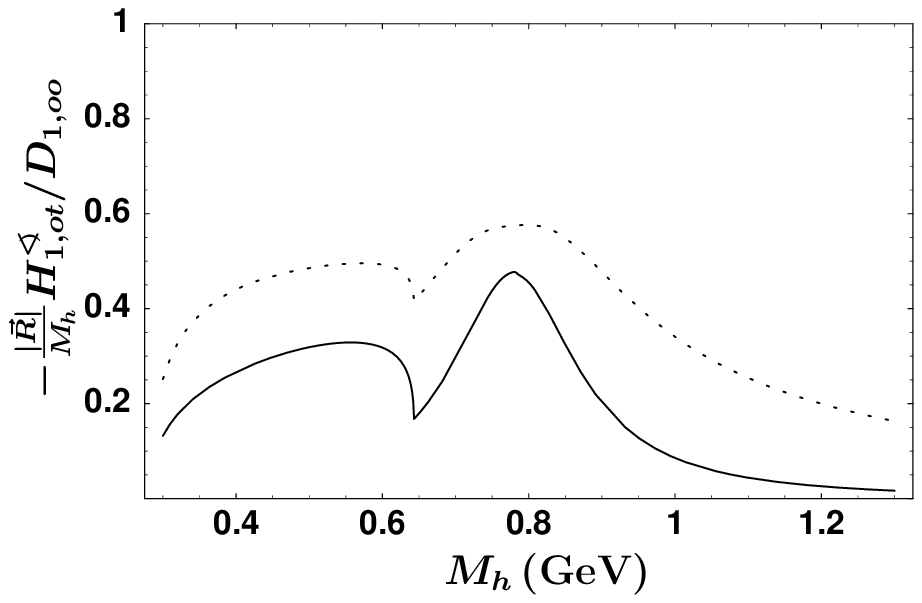}
&\hspace{1cm} &
\includegraphics[width=7cm]{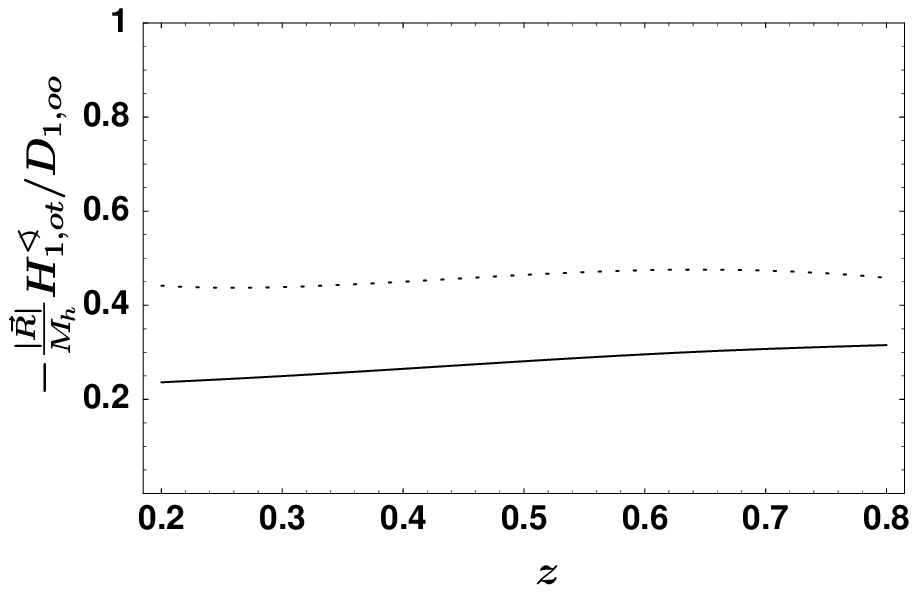}
\\
(a) & & (b)
\end{tabular}
\caption{Model prediction for the ratio $(-|\vec{R}|\, H_{1,ot}^{\open})/
  (M_H\,D_{1,oo})$: (a) as a function of $M_h$, (b) as a function of $z$. 
  The overall sign of $H_{1,ot}^{\open}$ cannot be predicted by the
  model and is chosen to have an agreement with the sign of preliminary HERMES
  measurements~\cite{vanderNat:2005yf}. 
  The dotted lines represent the positivity bounds of Eq.~(\ref{e:posh1ang}).}
\label{f:h1angle}
\end{figure}

In Fig.~\ref{f:h1angle} (a) we plot the ratio between 
$- |\vec{R}|/M_h \, H_{1,ot}^{\open}$
and $D_{1,oo}$, integrated separately 
over $0.2< z < 0.8$. In Fig.~\ref{f:h1angle} (b) we plot the same ratio
but with the two functions  multiplied by $2 M_h$ and integrated over $0.3\,\text{GeV} < M_h < 1.3\,\text{GeV}$. In the same
figures, the dotted lines represent 
the positivity
bound~\cite{Bacchetta:2002ux}
\begin{equation}
\frac{|\vec R|}{M_h}\,H_{1,ot}^{\open} \leq \sqrt{\frac{3}{8}\,D_{1,oo}^s
   \left(D_{1,oo}^p-\frac{1}{3}\,D_{1,ll}\right)}.
\label{e:posh1ang}
\end{equation} 

As is evident, there are two
main contributions:
\begin{itemize}
\item{the interference between channel 1 ($s$-wave background) and the
    imaginary part of 2 ($\rho$
    resonance), with a shape
    peaked at the $\rho$ mass, i.e.\ roughly proportional 
    to the imaginary part of the $\rho$ resonance in Eq.~(\ref{e:Fp});}
\item{the interference between channel 1 ($s$-wave background) and 4 ($\omega$
    resonance decaying into three pions), with a shape peaked at $M_h \approx 0.5$
    GeV, roughly proportional to
  the third (imaginary) term in Eq.~(\ref{e:Fp}).}
\end{itemize}
The two contributions have comparable size and are large. At this point, we
want to stress once more that our model assumptions imply that the
above channels can interfere in a complete way, since the spectators $X_1$,
$X_2$ and $X_4$ are the same. As already argued before, it is likely that only
a fraction of the $X_2$ and $X_1$ states interfere, and so does a (in general
different) fraction of  the $X_4$ and $X_1$ states. This could
decrease the sizes of the two ``peaks'' of Fig.~\ref{f:h1angle} (a) and
accordingly the
overall size of the curve in Fig.~\ref{f:h1angle} (b). This is beyond
the reach of our model in its present form, 
but could be a way to proceed when fitting data
related to $H_{1,ot}^{\open}$.

\begin{figure}
\begin{tabular}{ccc}
\includegraphics[width=7cm]{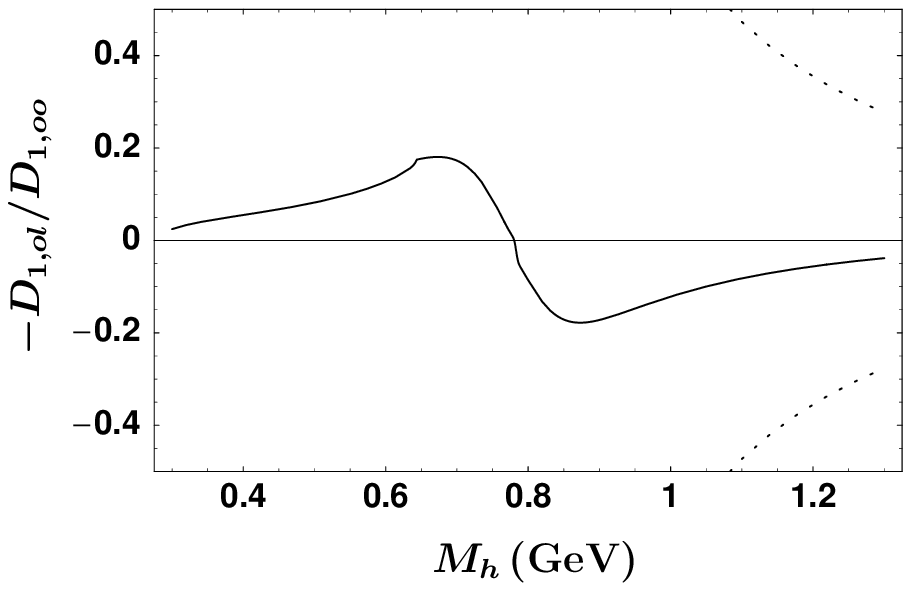}
&\hspace{1cm} &
\includegraphics[width=7cm]{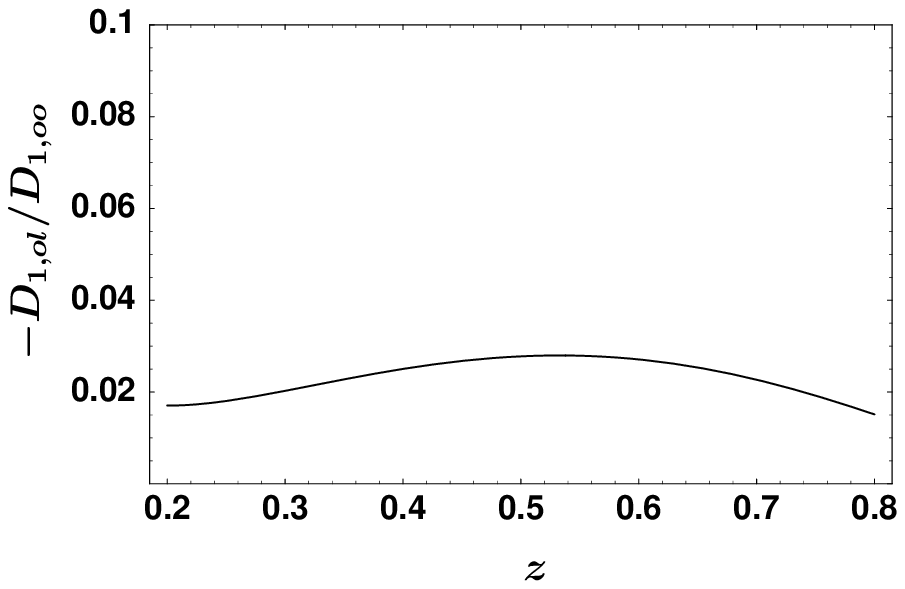}
\\
(a) & & (b)
\end{tabular}
\caption{Model prediction for the ratio $-D_{1,ol}/
  D_{1,oo}$: (a) as a function of $M_h$, (b) as a function of $z$. 
  The overall sign of $D_{1,ol}$ depends on the choice made on the sign of
  $H_{1,ot}^{\open}$.
  The dotted lines represent the positivity bounds of Eq.~(\ref{e:posd1sp}).}
\label{f:d1sp}
\end{figure}

In Fig.~\ref{f:d1sp} (a) we plot the ratio between 
$- D_{1,ol}$
and $D_{1,oo}$, integrated separately 
over $0.2< z < 0.8$. In Fig.~\ref{f:d1sp} (b) we plot the same ratio
but with the two functions multiplied by $2 M_h$ and integrated over $0.3\,\text{GeV} < M_h < 1.3\,\text{GeV}$. The dotted line
in Fig.~\ref{f:d1sp} (a) represents 
%In the same
%figures we plot also 
the positivity
bound~\cite{Bacchetta:2002ux} (in the
second plot the bound lies beyond the plot range)
\begin{equation}
D_{1,ol} \leq \sqrt{\frac{3}{4}\,D_{1,oo}^s \left(D_{1,oo}^p+\frac{2}{3}\,
   D_{1,ll}\right)}.
\label{e:posd1sp}
\end{equation} 

In this case, the function $D_{1,ol}$ receives basically 
only one contribution, namely 
from the interference between channel 1 ($s$-wave background) and the real
part of 2 ($\rho$ resonance). In fact, its shape has a sign change at the
$\rho$ mass and is roughly proportional to
the real part of the $\rho$ resonance in Eq.~(\ref{e:Fp}). Channel 3 is
negligible as in the previous case, while channel 4 plays no
role now because we assumed it to be purely imaginary.

\begin{figure}
\begin{tabular}{ccccc}
\includegraphics[width=5.5cm]{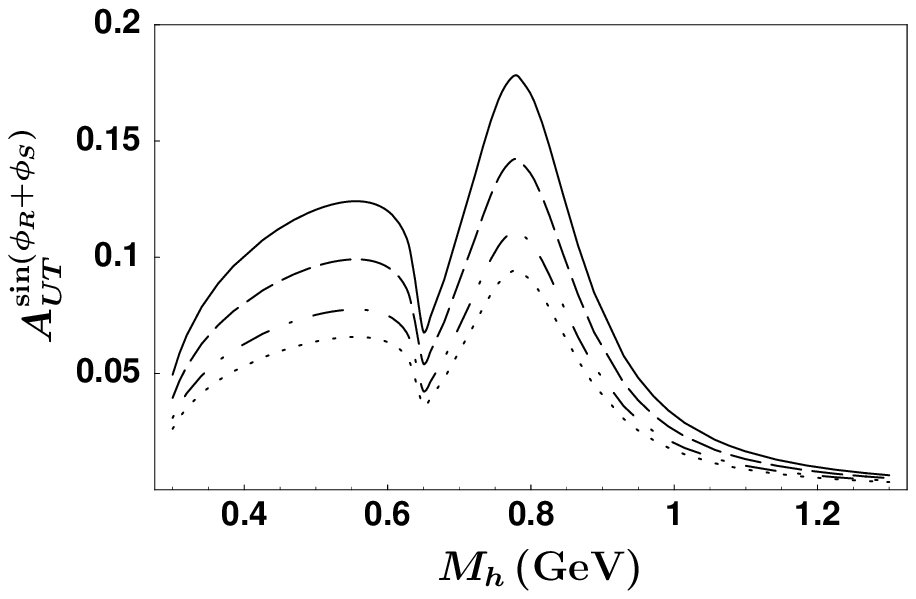}
&\hspace{0.3cm} &
\includegraphics[width=5.5cm]{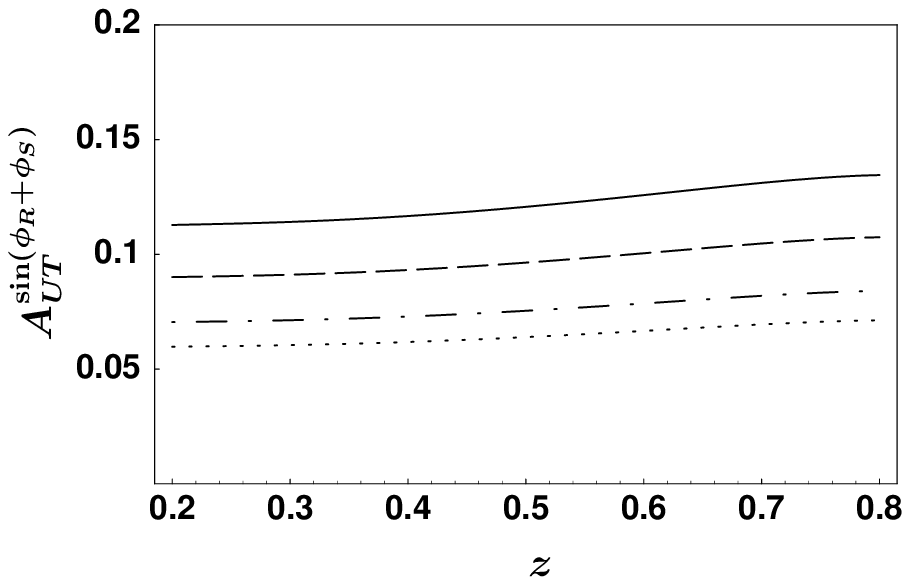}
&\hspace{0.3cm} &
\includegraphics[width=5.5cm]{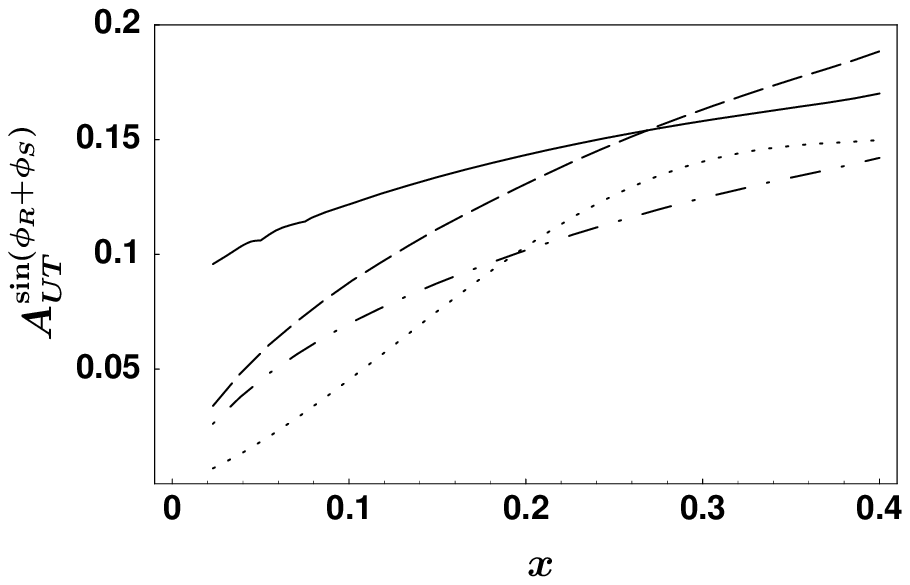}
\\
(a) & & (b) & & (c)
\end{tabular}
\caption{$A_{UT}^{\sin(\phi_R^{} + \phi_S^{})}$ moment defined in
  Eq.~(\ref{e:asy2}) at HERMES kinematics for a proton target: (a) as a
  function of $M_h$, (b) as a
  function of $z$, (c) as a
  function of $x$. The different lines correspond to different models of the
  transversity distribution function: dotted line from Ref.~\cite{Wakamatsu:2000fd}, dash-dotted line from Ref.~\cite{Korotkov:2001jx}, dashed
  line from Ref.~\cite{Schweitzer:2001sr}, solid line from Ref.~\cite{Soffer:2002tf}.}
\label{f:asy}
\end{figure}

Next we compute the asymmetry defined in Eq.~(\ref{e:asy2}), integrated over
all variables but one. In
Fig.~\ref{f:asy} (a) we plot
\begin{equation}
%\bigl|
A_{UT}^{\sin(\phi_R^{} + \phi_S^{})}(M_h)
%\bigr| 
=
c \; 
%\frac{|\vec{R}|}{M_h}
\frac{\int_{
0.2
%z_{\text{min}}
}^{
0.8
%z_{\text{max}}
}dz\;|\vec{R}|/M_h\;
%\bigl|
H_{1,ot}^{\open}(z,M_h^2)
%\bigr|
}{\int_{
0.2
%z_{\text{min}}
}^{
0.8
%z_{\text{max}}
}dz\;D_{1,oo}(z,M_h^2)},
\end{equation} 
and in Fig.~\ref{f:asy} (b) we plot
\begin{equation}
%\bigl|
A_{UT}^{\sin(\phi_R^{} + \phi_S^{})}(z)
%\bigr| 
=
c \; 
%\frac{|\vec{R}|}{M_h}
\frac{\int_{
0.5
%\, \text{GeV}
%M_{h\, \text{min}}
}^{
1
%\, \text{GeV}
%M_{h\, \text{max}}
}dM_h\;2M_h\;|\vec{R}|/M_h\;
%\bigl|
H_{1,ot}^{\open}(z,M_h^2)
%\bigr|
}{\int_{
0.5
%\, \text{GeV}
%M_{h\, \text{min}}
}^{
1
%\, \text{GeV}
%M_{h\, \text{max}}
}dM_h\;2M_h\;D_{1,oo}(z,M_h^2)},
\end{equation} 
where
\begin{equation}
\label{e:c}
c=\frac{\pi}{4}\,\frac{\int_{
0.1
%y_{\text{min}}
}^{
0.85
%y_{\text{max}}
}dy\;
% (1-y)/y^2
    \int_{x_{\text{min}}}^{
%0.4
x_{\text{max}}
} dx\;\frac{1-y-y^2\,\gamma^2/4}{x\,y^2\,(1+\gamma^2)}
\;
\bigr[4\,\bigr(h_1^u(x)-h_1^{\bar{u}}(x)\bigl)+\bigr(h_1^{\bar{d}}(x) -h_1^d(x)\bigl)\bigl]}
{\int_{
0.1
%y_{\text{min}}
}^{
0.85
%y_{\text{max}}
} dy\;
%(1-y+y^2/2)/y^2
    \int_{x_{\text{min}}}^{
%0.4
x_{\text{max}}
}
      dx\;\frac{1-y+y^2/2 + y^2\,\gamma^2/4}{x\,y^2\,(1+\gamma^2)}
\;
\bigr[4\,\bigr(f_1^u(x)+f_1^{\bar{u}}(x)\bigl)+\bigr(f_1^{\bar{d}}(x) +f_1^d(x)\bigl)\bigl]}.
\end{equation} 
We neglected strange quark contributions.
The fact that the $d$ and $\bar{u}$ transversity distributions enter with an
opposite sign is due to the symmetry properties of the fragmentation
functions. As already discussed in Sec.~\ref{s:model}, 
the fragmentation function $H_{1,ot}^{\open}$ is the same for all
quarks, but the sign of $\sin(\phi_R^{} + \phi_S^{})$ changes for $d$ and
$\bar{u}$. 

In Fig.~\ref{f:asy} (c) we plot
\begin{equation}
%\bigl|
A_{UT}^{\sin(\phi_R^{} + \phi_S^{})}(x)
%\bigr| 
=
c' \; 
\frac{\int_{
y_{\text{min}}
}^{
0.85
%y_{\text{max}}
}dy\;
% (1-y)/y^2
\frac{1-y-y^2\,\gamma^2/4}{y^2\,(1+\gamma^2)}
\;
\bigr[4\,\bigr(h_1^u(x)-h_1^{\bar{u}}(x)\bigl)+\bigr(h_1^{\bar{d}}(x) -h_1^d(x)\bigl)\bigl]}
{\int_{
y_{\text{min}}
}^{
0.85
%y_{\text{max}}
} dy\;
%(1-y+y^2/2)/y^2
\frac{1-y+y^2/2 + y^2\,\gamma^2/4}{y^2\,(1+\gamma^2)}
\;
\bigr[4\,\bigr(f_1^u(x)+f_1^{\bar{u}}(x)\bigl)+\bigr(f_1^{\bar{d}}(x)
+f_1^d(x)\bigl)\bigl]},
\end{equation} 
where 
\begin{equation}
c'
=
%\frac{|\vec{R}|}{M_h}
\frac{\pi}{4}\,\frac{\int_{
0.2
%z_{\text{min}}
}^{
0.8
%z_{\text{max}}
}dz\;
\int_{
0.5
%\, \text{GeV}
%M_{h\, \text{min}}
}^{
1
%\, \text{GeV}
%M_{h\, \text{max}}
}dM_h\;2M_h\;|\vec{R}|/M_h\;
%\bigl|
H_{1,ot}^{\open}(z,M_h^2)
%\bigr|
}{
\int_{
0.2
%z_{\text{min}}
}^{
0.8
%z_{\text{max}}
}dz\;
\int_{
0.5
%\, \text{GeV}
%M_{h\, \text{min}}
}^{
1
%\, \text{GeV}
%M_{h\, \text{max}}
}dM_h\;2M_h\;D_{1,oo}(z,M_h^2)}.
\end{equation} 

%\begin{figure}
%\begin{tabular}{ccccc}
%\includegraphics[width=5.5cm]{asymh}
%&\hspace{0.5cm} &
%\includegraphics[width=5.5cm]{asyz}
%&\hspace{0.5cm} &
%\includegraphics[width=5.5cm]{asyx}
%\\
%(a) & & (b) & & (c)
%\end{tabular}
%\caption{$A_{UT}^{\sin(\phi_R^{} + \phi_S^{})}$ amplitude (a) as a
%  function of $M_h$, (b) as a
%  function of $z$, (c) as a
%  function of $x$, 
%.}
%\label{f:asy}
%\end{figure}

The choices of the integrations boundaries for $x$, $y$, $z$ and $M_h$ 
are inspired by the HERMES
cuts~\cite{vanderNat:2005yf}. We took ($s= 56.2$ GeV$^2$)
\begin{align}   
x_{\text{min}} &= \text{Max}\bigl[0.023,
Q^2_{\text{min}}/\bigl(y(s-M^2)\bigr)\bigr],
\\
x_{\text{max}} &=
\text{Min}\bigl[0.4,1-(W^2_{\text{min}}-M^2)/\bigl(y(s-M^2)\bigr)\bigr],
\\
y_{\text{min}} &= \text{Max}\bigl[0.1,
Q^2_{\text{min}}/\bigl(x(s-M^2)\bigr),(W^2_{\text{min}}-M^2)/\bigl((1-x)(s-M^2)\bigr)\bigr].
\end{align}  
%Note that expressing the asymmetry as a product of a
%coefficient times the integral of the fragmentation functions is allowed only
%if the pion momentum cuts are neglected.
For the unpolarized parton distribution functions we take the parameterization
of Ref.~\cite{Gluck:1998xa}. For the transversity distribution function, we
take the estimates of
Refs.~\cite{Wakamatsu:2000fd,Korotkov:2001jx,Schweitzer:2001sr,Soffer:2002tf}.
%Even assuming the smallest estimate for the transversity distribution, 
The sign of the preliminary data indicates that the
$s$-wave and $p$-wave couplings should have opposite signs and thus
$H_{1,ot}^{\open}$ should be negative.
The asymmetry obtained
from our model appears to overestimate the preliminary HERMES
data~\cite{vanderNat:2005yf} by about a factor 3-4. This probably indicates that the model overestimates in general the effect of
interferences.
Apart from the overall normalization, 
the height of the bump around $M_h
\approx 0.5$ GeV seems to be too big relative to the $\rho$ peak, which 
is probably due to the fact that   
not all the $\pi^+ \pi^-$ pairs in channel 4 should be considered
in $p$ wave. 
%The invariant mass and $z$ shapes could well be compatible with
%the preliminary data.  
However, in order to make more conclusive statements it is necessary to wait
for HERMES final data. Obviously, it would be
better to compare our model with an observable where $H_{1,ot}^{\open}$ can
be isolated, e.g., in $e^+e^-$ annihilation at BELLE~\cite{Hasuko:2003ay}. 

\begin{figure}
\begin{tabular}{ccccc}
\includegraphics[width=5.5cm]{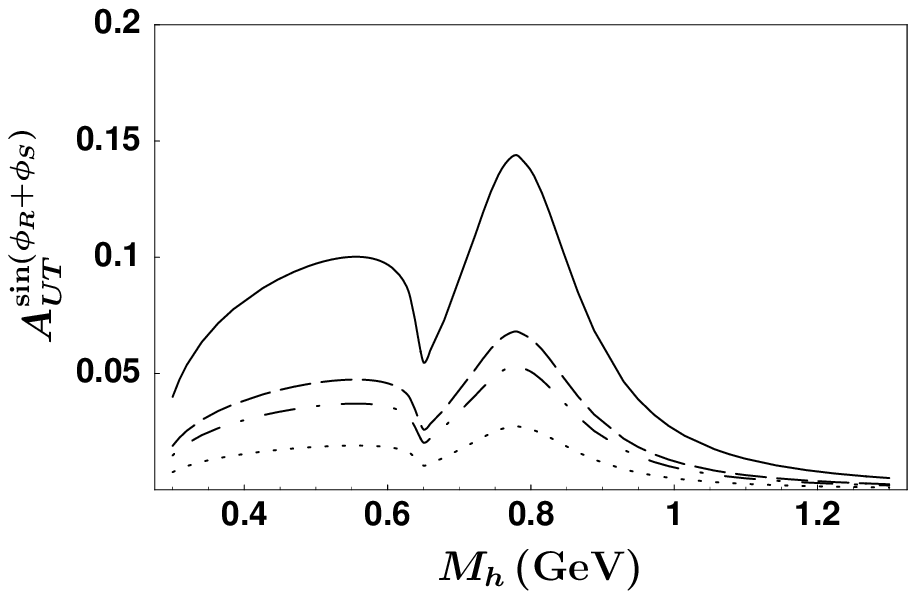}
&\hspace{0.3cm} &
\includegraphics[width=5.5cm]{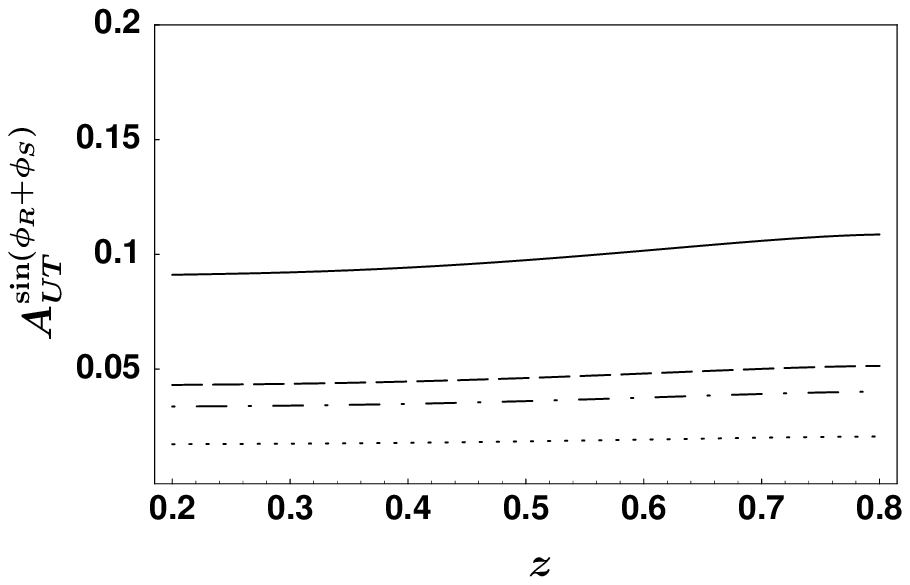}
&\hspace{0.3cm} &
\includegraphics[width=5.5cm]{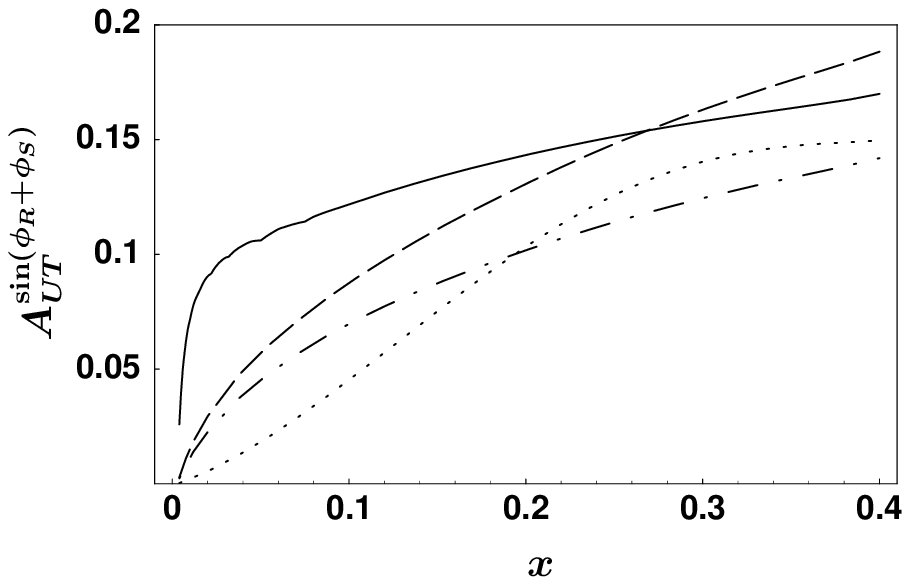}
\\
(a) & & (b) & & (c)
\end{tabular}
\caption{$A_{UT}^{\sin(\phi_R^{} + \phi_S^{})}$ moment defined in
  Eq.~(\ref{e:asy2}) at COMPASS kinematics for a proton target: (a) as a
  function of $M_h$, (b) as a
  function of $z$, (c) as a
  function of $x$. The different lines correspond to different models of the
  transversity distribution function: dotted line from Ref.~\cite{Wakamatsu:2000fd}, dash-dotted line from Ref.~\cite{Korotkov:2001jx}, dashed
  line from Ref.~\cite{Schweitzer:2001sr}, solid line from Ref.~\cite{Soffer:2002tf}.}
\label{f:asy_compprot}
\end{figure}

In
Fig.~\ref{f:asy_compprot} we plot the same asymmetry as before, but for 
the kinematics of the COMPASS experiment. 
We assumed the same cuts as
before  and change only the value of $s$.
The size of the $M_h$- and $z$-dependent
asymmetries is smaller than at HERMES. This is due to the sensitivity
of COMPASS to lower values of $x$, where models predict transversity 
to be small, while the unpolarized
distribution functions are big.
Due to the same reason, there is a much larger difference among the
models, as they differ substantially at low $x$. 
The asymmetries could be enhanced if the low-$x$ region is
excluded from the integration.

\begin{figure}
\begin{tabular}{ccccc}
\includegraphics[width=5.5cm]{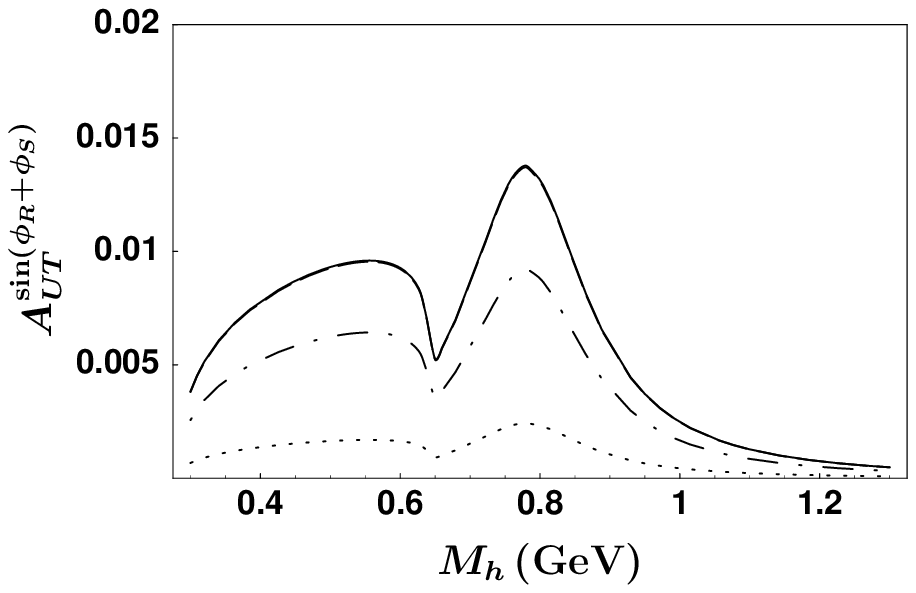}
&\hspace{0.3cm} &
\includegraphics[width=5.5cm]{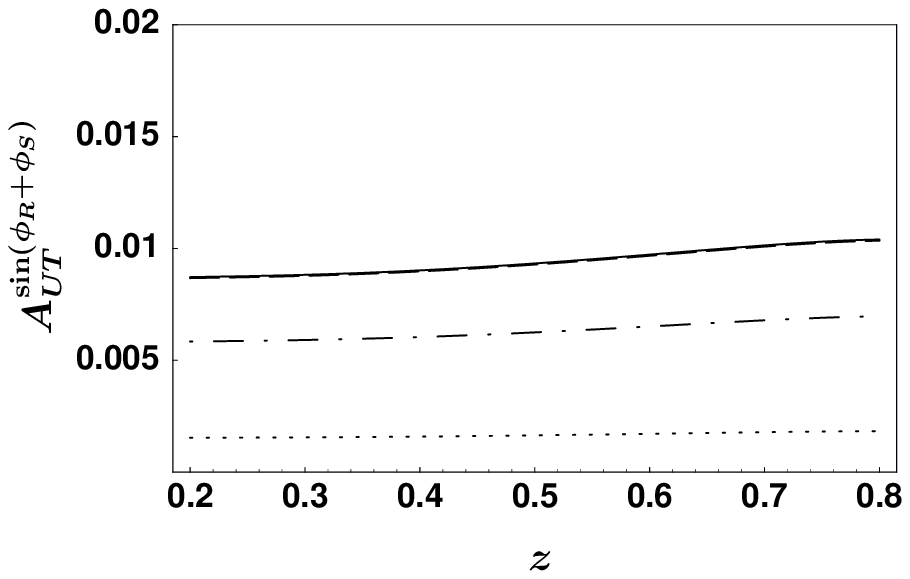}
&\hspace{0.3cm} &
\includegraphics[width=5.5cm]{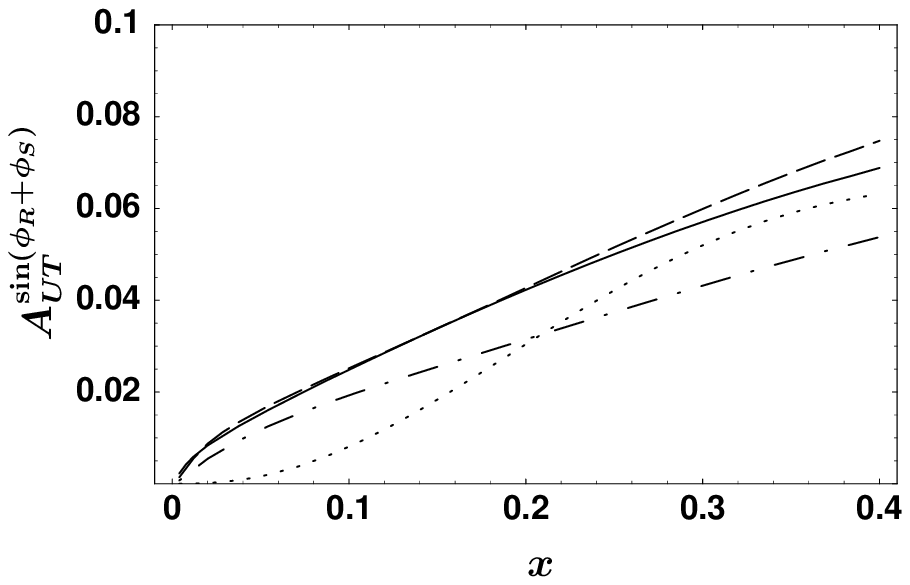}
\\
(a) & & (b) & & (c)
\end{tabular}
\caption{$A_{UT}^{\sin(\phi_R^{} + \phi_S^{})}$ moment defined in
  Eq.~(\ref{e:asy2}) at COMPASS kinematics for a deuteron target: (a) as a
  function of $M_h$, (b) as a
  function of $z$, (c) as a
  function of $x$. The different lines correspond to different models of the
  transversity distribution function: dotted line from Ref.~\cite{Wakamatsu:2000fd}, dash-dotted line from Ref.~\cite{Korotkov:2001jx}, dashed
  line from Ref.~\cite{Schweitzer:2001sr}, solid line from Ref.~\cite{Soffer:2002tf}.}
\label{f:asy_compdeu}
\end{figure}

The COMPASS collaboration has also presented preliminary data of the above 
asymmetry for a
deuteron target~\cite{Martin:2006}. We plot our prediction in
Fig.~\ref{f:asy_compdeu}.\footnote{Note that the 
        preliminary measurements of COMPASS correspond
  to $-A_{UT}^{\sin(\phi_R^{} + \phi_S^{})}$.} 
The different isospin structure of the target, combined with that of
the fragmentation functions in our 
model, decreases the asymmetry. The $x$-dependent asymmetry is less than half
of that for the proton target, while 
the $M_h$- and $z$-dependent asymmetries are
about 10 times smaller than for the proton target.

%%%%%%%%%%%%%%%%%%%%%%%%%%%%%%%%%%%%%%%%%%%%%%%%%%%%%%%%%%%%%%%%%%%%%%%%%%%%%
\section{Conclusions}
\label{s:conc}

In this paper we presented a model for the process $q\to \pi^+ \pi^- X$
at invariant mass
$M_h\lesssim 1.3$ GeV. We used a ``spectator'' model, where the sum over
all possible intermediate states $X$ is replaced by an effective on-shell
state.
Using this model we calculated the fragmentation functions that
can be defined at leading twist when considering only relative $s$ and $p$
waves of the pion pair~\cite{Bacchetta:2002ux}. 
We obtained nonzero results for four out of five of them.

We fixed the values of the parameters of the model by comparing the
unpolarized fragmentation function $D_{1,oo}$ with the
output of the PYTHIA event generator~\cite{Sjostrand:2000wi} tuned for 
HERMES~\cite{Liebing:2004us}. 
The main characteristics of the $M_h$ and $z$ shapes of 
$D_{1,oo}$ are qualitatively well described.

We made predictions for the fragmentation functions $D_{1,ll}$,  $D_{1,ol}$,
and $H_{1,ot}^{\open}$. The first one is a pure $p$-wave function, it is found
to be positive, about 50\% of the unpolarized fragmentation function and with
peaks at the $\rho$ mass and at around $M_h \approx 0.5$ GeV, where 
the $\omega$ decaying into three pions gives a large contribution.

The function $D_{1,ol}$ arises from the interference between $s$ and $p$
wave. Since in our model we assumed the $s$ wave to be purely real, this
function turns out to be proportional to the real part of the $p$ wave and in
particular displays a sign change at the $\rho$ mass. The size of the function
is small, in particular when integrated over the invariant mass, due to the
sign change. Our model cannot predict the overall sign of the function.

The function $H_{1,ot}^{\open}$ also arises from the interference between 
$s$ and $p$ waves, but is proportional to the imaginary part of the $p$
wave, i.e., it has peaks at the $\rho$ mass and at around $M_h \approx 0.5$
GeV, due to the contribution of the $\omega \to 3\pi$ channel. 
Its size is about 30\% of the unpolarized fragmentation function. Our
model cannot predict the overall sign of the function.

The function $H_{1,ot}^{\open}$ is of particular interest because in
two-hadron-inclusive deep inelastic scattering off transversely polarized
targets it gives rise to a single-spin
asymmetry in combination with the transversity distribution
function. Therefore, it could be used as an analyzer for this so far unknown
distribution function.
We estimated this single-spin asymmetry at HERMES kinematics 
using four different models for
the transversity distribution function. We found the asymmetry to be of the
order of 
10\% on average. 
The sign of the preliminary HERMES
measurements suggests that $H_{1,ot}^{\open}$ should be negative. 
The measurement indicates that the asymmetry in our model 
is about 3-4 times bigger than the data. This probably means that our model
overestimates the effects of interferences. However,
final experimental results are needed to make more reliable comparisons.

For COMPASS kinematics, the enhanced sensitivity to the portion of phase space
at very low $x$ induces a reduction in the spin asymmetry with respect to
HERMES, which can largely differ depending on the model for transversity. For
the deuteron target, the particular isospin structure, combined with that of
the fragmentation functions in our 
model, induces a further reduction such that the resulting asymmetry is
much smaller than for the proton, in agreement with preliminary data of the
COMPASS collaboration.

%%%%%%%%%%%%%%%%%%%%%%%%%%%%%%%%%%%%%%%%%%%%%%%%%%%%%%%%%%%%%%%%%%%%%%%%%%%%%

\begin{acknowledgments}
Useful discussions with P.~van der Nat, C.~A.~Miller, G.~Schnell,
E.~C.~Aschenauer, are thankfully acknowledged.  We are particularly grateful
to the HERMES Collaboration 
for providing us with the PYTHIA output. We are grateful to M.~Stratmann,
V.~Korotkov, P.~Schweitzer and  M.~Wakamatsu for making their predictions for the transversity
distribution function available. This work is partially supported by 
the European Integrated Infrastructure Initiative in Hadron Physics project 
under the contract number RII3-CT-2004-506078.
\end{acknowledgments}

%%%%%%%%%%%%%%%%%%%%%%%%%%%%%%%%%%%%%%%%%%%%%%%%%%%%%%%%%%%%%%%%%%%%%%%%%%%%%

\bibliographystyle{apsrev}
\bibliography{mybiblio}

\begin{thebibliography}{71}
\expandafter\ifx\csname natexlab\endcsname\relax\def\natexlab#1{#1}\fi
\expandafter\ifx\csname bibnamefont\endcsname\relax
  \def\bibnamefont#1{#1}\fi
\expandafter\ifx\csname bibfnamefont\endcsname\relax
  \def\bibfnamefont#1{#1}\fi
\expandafter\ifx\csname citenamefont\endcsname\relax
  \def\citenamefont#1{#1}\fi
\expandafter\ifx\csname url\endcsname\relax
  \def\url#1{\texttt{#1}}\fi
\expandafter\ifx\csname urlprefix\endcsname\relax\def\urlprefix{URL }\fi
\providecommand{\bibinfo}[2]{#2}
\providecommand{\eprint}[2][]{\url{#2}}

\bibitem[{\citenamefont{Konishi et~al.}(1978)\citenamefont{Konishi, Ukawa, and
  Veneziano}}]{Konishi:1978yx}
\bibinfo{author}{\bibfnamefont{K.}~\bibnamefont{Konishi}},
  \bibinfo{author}{\bibfnamefont{A.}~\bibnamefont{Ukawa}}, \bibnamefont{and}
  \bibinfo{author}{\bibfnamefont{G.}~\bibnamefont{Veneziano}},
  \bibinfo{journal}{Phys. Lett.} \textbf{\bibinfo{volume}{B78}},
  \bibinfo{pages}{243} (\bibinfo{year}{1978}).

\bibitem[{\citenamefont{Vendramin}(1981)}]{Vendramin:1981te}
\bibinfo{author}{\bibfnamefont{I.}~\bibnamefont{Vendramin}},
  \bibinfo{journal}{Nuovo Cim.} \textbf{\bibinfo{volume}{A66}},
  \bibinfo{pages}{339} (\bibinfo{year}{1981}).

\bibitem[{\citenamefont{Sukhatme and Lassila}(1980)}]{Sukhatme:1980vs}
\bibinfo{author}{\bibfnamefont{U.~P.} \bibnamefont{Sukhatme}} \bibnamefont{and}
  \bibinfo{author}{\bibfnamefont{K.~E.} \bibnamefont{Lassila}},
  \bibinfo{journal}{Phys. Rev.} \textbf{\bibinfo{volume}{D22}},
  \bibinfo{pages}{1184} (\bibinfo{year}{1980}).

\bibitem[{\citenamefont{de~Florian and Vanni}(2004)}]{deFlorian:2003cg}
\bibinfo{author}{\bibfnamefont{D.}~\bibnamefont{de~Florian}} \bibnamefont{and}
  \bibinfo{author}{\bibfnamefont{L.}~\bibnamefont{Vanni}},
  \bibinfo{journal}{Phys. Lett.} \textbf{\bibinfo{volume}{B578}},
  \bibinfo{pages}{139} (\bibinfo{year}{2004}), \eprint{hep-ph/0310196}.

\bibitem[{\citenamefont{Majumder and Wang}(2004)}]{Majumder:2004wh}
\bibinfo{author}{\bibfnamefont{A.}~\bibnamefont{Majumder}} \bibnamefont{and}
  \bibinfo{author}{\bibfnamefont{X.-N.} \bibnamefont{Wang}},
  \bibinfo{journal}{Phys. Rev.} \textbf{\bibinfo{volume}{D70}},
  \bibinfo{pages}{014007} (\bibinfo{year}{2004}), \eprint{hep-ph/0402245}.

\bibitem[{\citenamefont{Majumder and Wang}(2005)}]{Majumder:2004br}
\bibinfo{author}{\bibfnamefont{A.}~\bibnamefont{Majumder}} \bibnamefont{and}
  \bibinfo{author}{\bibfnamefont{X.-N.} \bibnamefont{Wang}},
  \bibinfo{journal}{Phys. Rev.} \textbf{\bibinfo{volume}{D72}},
  \bibinfo{pages}{034007} (\bibinfo{year}{2005}), \eprint{hep-ph/0411174}.

\bibitem[{\citenamefont{Acton et~al.}(1992)}]{Acton:1992sa}
\bibinfo{author}{\bibfnamefont{P.~D.} \bibnamefont{Acton}} \bibnamefont{et~al.}
  (\bibinfo{collaboration}{OPAL}), \bibinfo{journal}{Z. Phys.}
  \textbf{\bibinfo{volume}{C56}}, \bibinfo{pages}{521} (\bibinfo{year}{1992}).

\bibitem[{\citenamefont{Abreu et~al.}(1993)}]{Abreu:1992xx}
\bibinfo{author}{\bibfnamefont{P.}~\bibnamefont{Abreu}} \bibnamefont{et~al.}
  (\bibinfo{collaboration}{DELPHI}), \bibinfo{journal}{Phys. Lett.}
  \textbf{\bibinfo{volume}{B298}}, \bibinfo{pages}{236} (\bibinfo{year}{1993}).

\bibitem[{\citenamefont{Buskulic et~al.}(1996)}]{Buskulic:1995gm}
\bibinfo{author}{\bibfnamefont{D.}~\bibnamefont{Buskulic}} \bibnamefont{et~al.}
  (\bibinfo{collaboration}{ALEPH}), \bibinfo{journal}{Z. Phys.}
  \textbf{\bibinfo{volume}{C69}}, \bibinfo{pages}{379} (\bibinfo{year}{1996}).

\bibitem[{\citenamefont{Cohen et~al.}(1982)}]{Cohen:1982zg}
\bibinfo{author}{\bibfnamefont{I.}~\bibnamefont{Cohen}} \bibnamefont{et~al.},
  \bibinfo{journal}{Phys. Rev.} \textbf{\bibinfo{volume}{D25}},
  \bibinfo{pages}{634} (\bibinfo{year}{1982}).

\bibitem[{\citenamefont{Aubert et~al.}(1983)}]{Aubert:1983un}
\bibinfo{author}{\bibfnamefont{J.~J.} \bibnamefont{Aubert}}
  \bibnamefont{et~al.} (\bibinfo{collaboration}{European Muon}),
  \bibinfo{journal}{Phys. Lett.} \textbf{\bibinfo{volume}{B133}},
  \bibinfo{pages}{370} (\bibinfo{year}{1983}).

\bibitem[{\citenamefont{Arneodo et~al.}(1986)}]{Arneodo:1986tc}
\bibinfo{author}{\bibfnamefont{M.}~\bibnamefont{Arneodo}} \bibnamefont{et~al.}
  (\bibinfo{collaboration}{European Muon}), \bibinfo{journal}{Z. Phys.}
  \textbf{\bibinfo{volume}{C33}}, \bibinfo{pages}{167} (\bibinfo{year}{1986}).

\bibitem[{\citenamefont{Blobel et~al.}(1974)}]{Blobel:1973wr}
\bibinfo{author}{\bibfnamefont{V.}~\bibnamefont{Blobel}} \bibnamefont{et~al.}
  (\bibinfo{collaboration}{Bonn-Hamburg-Munich}), \bibinfo{journal}{Phys.
  Lett.} \textbf{\bibinfo{volume}{B48}}, \bibinfo{pages}{73}
  (\bibinfo{year}{1974}).

\bibitem[{\citenamefont{Aguilar-Benitez et~al.}(1991)}]{Aguilar-Benitez:1991yy}
\bibinfo{author}{\bibfnamefont{M.}~\bibnamefont{Aguilar-Benitez}}
  \bibnamefont{et~al.}, \bibinfo{journal}{Z. Phys.}
  \textbf{\bibinfo{volume}{C50}}, \bibinfo{pages}{405} (\bibinfo{year}{1991}).

\bibitem[{\citenamefont{Adams et~al.}(2004)}]{Adams:2003cc}
\bibinfo{author}{\bibfnamefont{J.}~\bibnamefont{Adams}} \bibnamefont{et~al.}
  (\bibinfo{collaboration}{STAR}), \bibinfo{journal}{Phys. Rev. Lett.}
  \textbf{\bibinfo{volume}{92}}, \bibinfo{pages}{092301}
  (\bibinfo{year}{2004}), \eprint{nucl-ex/0307023}.

\bibitem[{\citenamefont{Fachini}(2004)}]{Fachini:2004jx}
\bibinfo{author}{\bibfnamefont{P.}~\bibnamefont{Fachini}}, \bibinfo{journal}{J.
  Phys.} \textbf{\bibinfo{volume}{G30}}, \bibinfo{pages}{S735}
  (\bibinfo{year}{2004}), \eprint{nucl-ex/0403026}.

\bibitem[{\citenamefont{Majumder}(2005)}]{Majumder:2005vs}
\bibinfo{author}{\bibfnamefont{A.}~\bibnamefont{Majumder}},
  \bibinfo{journal}{J. Phys. Conf. Ser.} \textbf{\bibinfo{volume}{9}},
  \bibinfo{pages}{294} (\bibinfo{year}{2005}), \eprint{nucl-th/0501029}.

\bibitem[{\citenamefont{Efremov et~al.}(1992)\citenamefont{Efremov, Mankiewicz,
  and Tornqvist}}]{Efremov:1992pe}
\bibinfo{author}{\bibfnamefont{A.~V.} \bibnamefont{Efremov}},
  \bibinfo{author}{\bibfnamefont{L.}~\bibnamefont{Mankiewicz}},
  \bibnamefont{and} \bibinfo{author}{\bibfnamefont{N.~A.}
  \bibnamefont{Tornqvist}}, \bibinfo{journal}{Phys. Lett.}
  \textbf{\bibinfo{volume}{B284}}, \bibinfo{pages}{394} (\bibinfo{year}{1992}).

\bibitem[{\citenamefont{Collins et~al.}(1994)\citenamefont{Collins, Heppelmann,
  and Ladinsky}}]{Collins:1994kq}
\bibinfo{author}{\bibfnamefont{J.~C.} \bibnamefont{Collins}},
  \bibinfo{author}{\bibfnamefont{S.~F.} \bibnamefont{Heppelmann}},
  \bibnamefont{and} \bibinfo{author}{\bibfnamefont{G.~A.}
  \bibnamefont{Ladinsky}}, \bibinfo{journal}{Nucl. Phys.}
  \textbf{\bibinfo{volume}{B420}}, \bibinfo{pages}{565} (\bibinfo{year}{1994}),
  \eprint[http://arXiv.org/abs]{hep-ph/9305309}.

\bibitem[{\citenamefont{Collins and Ladinsky}(1994)}]{Collins:1994ax}
\bibinfo{author}{\bibfnamefont{J.~C.} \bibnamefont{Collins}} \bibnamefont{and}
  \bibinfo{author}{\bibfnamefont{G.~A.} \bibnamefont{Ladinsky}}
  (\bibinfo{year}{1994}), \eprint[http://arXiv.org/abs]{hep-ph/9411444}.

\bibitem[{\citenamefont{Jaffe et~al.}(1998)\citenamefont{Jaffe, Jin, and
  Tang}}]{Jaffe:1998hf}
\bibinfo{author}{\bibfnamefont{R.~L.} \bibnamefont{Jaffe}},
  \bibinfo{author}{\bibfnamefont{X.}~\bibnamefont{Jin}}, \bibnamefont{and}
  \bibinfo{author}{\bibfnamefont{J.}~\bibnamefont{Tang}},
  \bibinfo{journal}{Phys. Rev. Lett.} \textbf{\bibinfo{volume}{80}},
  \bibinfo{pages}{1166} (\bibinfo{year}{1998}),
  \eprint[http://arXiv.org/abs]{hep-ph/9709322}.

\bibitem[{\citenamefont{Artru and Collins}(1996)}]{Artru:1996zu}
\bibinfo{author}{\bibfnamefont{X.}~\bibnamefont{Artru}} \bibnamefont{and}
  \bibinfo{author}{\bibfnamefont{J.~C.} \bibnamefont{Collins}},
  \bibinfo{journal}{Z. Phys.} \textbf{\bibinfo{volume}{C69}},
  \bibinfo{pages}{277} (\bibinfo{year}{1996}),
  \eprint[http://arXiv.org/abs]{hep-ph/9504220}.

\bibitem[{\citenamefont{Efremov and Teryaev}(1982)}]{Efremov:1982sh}
\bibinfo{author}{\bibfnamefont{A.~V.} \bibnamefont{Efremov}} \bibnamefont{and}
  \bibinfo{author}{\bibfnamefont{O.~V.} \bibnamefont{Teryaev}},
  \bibinfo{journal}{Sov. J. Nucl. Phys.} \textbf{\bibinfo{volume}{36}},
  \bibinfo{pages}{140} (\bibinfo{year}{1982}).

\bibitem[{\citenamefont{Ji}(1994)}]{Ji:1994vw}
\bibinfo{author}{\bibfnamefont{X.}~\bibnamefont{Ji}}, \bibinfo{journal}{Phys.
  Rev.} \textbf{\bibinfo{volume}{D49}}, \bibinfo{pages}{114}
  (\bibinfo{year}{1994}), \eprint[http://arXiv.org/abs]{hep-ph/9307235}.

\bibitem[{\citenamefont{Anselmino et~al.}(1999)\citenamefont{Anselmino,
  Bertini, Caruso, Murgia, and Quintairos}}]{Anselmino:1999cg}
\bibinfo{author}{\bibfnamefont{M.}~\bibnamefont{Anselmino}},
  \bibinfo{author}{\bibfnamefont{M.}~\bibnamefont{Bertini}},
  \bibinfo{author}{\bibfnamefont{F.}~\bibnamefont{Caruso}},
  \bibinfo{author}{\bibfnamefont{F.}~\bibnamefont{Murgia}}, \bibnamefont{and}
  \bibinfo{author}{\bibfnamefont{P.}~\bibnamefont{Quintairos}},
  \bibinfo{journal}{Eur. Phys. J.} \textbf{\bibinfo{volume}{C11}},
  \bibinfo{pages}{529} (\bibinfo{year}{1999}), \eprint{hep-ph/9904205}.

\bibitem[{\citenamefont{Bacchetta and Mulders}(2000)}]{Bacchetta:2000jk}
\bibinfo{author}{\bibfnamefont{A.}~\bibnamefont{Bacchetta}} \bibnamefont{and}
  \bibinfo{author}{\bibfnamefont{P.~J.} \bibnamefont{Mulders}},
  \bibinfo{journal}{Phys. Rev.} \textbf{\bibinfo{volume}{D62}},
  \bibinfo{pages}{114004} (\bibinfo{year}{2000}),
  \eprint[http://arXiv.org/abs]{hep-ph/0007120}.

\bibitem[{\citenamefont{Bianconi
  et~al.}(2000{\natexlab{a}})\citenamefont{Bianconi, Boffi, Jakob, and
  Radici}}]{Bianconi:1999cd}
\bibinfo{author}{\bibfnamefont{A.}~\bibnamefont{Bianconi}},
  \bibinfo{author}{\bibfnamefont{S.}~\bibnamefont{Boffi}},
  \bibinfo{author}{\bibfnamefont{R.}~\bibnamefont{Jakob}}, \bibnamefont{and}
  \bibinfo{author}{\bibfnamefont{M.}~\bibnamefont{Radici}},
  \bibinfo{journal}{Phys. Rev.} \textbf{\bibinfo{volume}{D62}},
  \bibinfo{pages}{034008} (\bibinfo{year}{2000}{\natexlab{a}}),
  \eprint[http://arXiv.org/abs]{hep-ph/9907475}.

\bibitem[{\citenamefont{Bacchetta and
  Radici}(2004{\natexlab{a}})}]{Bacchetta:2003vn}
\bibinfo{author}{\bibfnamefont{A.}~\bibnamefont{Bacchetta}} \bibnamefont{and}
  \bibinfo{author}{\bibfnamefont{M.}~\bibnamefont{Radici}},
  \bibinfo{journal}{Phys. Rev.} \textbf{\bibinfo{volume}{D69}},
  \bibinfo{pages}{074026} (\bibinfo{year}{2004}{\natexlab{a}}),
  \eprint{hep-ph/0311173}.

\bibitem[{\citenamefont{Abreu et~al.}(1997)}]{Abreu:1997wd}
\bibinfo{author}{\bibfnamefont{P.}~\bibnamefont{Abreu}} \bibnamefont{et~al.}
  (\bibinfo{collaboration}{DELPHI}), \bibinfo{journal}{Phys. Lett.}
  \textbf{\bibinfo{volume}{B406}}, \bibinfo{pages}{271} (\bibinfo{year}{1997}).

\bibitem[{\citenamefont{Abbiendi et~al.}(2000)}]{Abbiendi:1999bz}
\bibinfo{author}{\bibfnamefont{G.}~\bibnamefont{Abbiendi}} \bibnamefont{et~al.}
  (\bibinfo{collaboration}{OPAL}), \bibinfo{journal}{Eur. Phys. J.}
  \textbf{\bibinfo{volume}{C16}}, \bibinfo{pages}{61} (\bibinfo{year}{2000}),
  \eprint[http://arXiv.org/abs]{hep-ex/9906043}.

\bibitem[{\citenamefont{Abe et~al.}(1995)}]{Abe:1994bk}
\bibinfo{author}{\bibfnamefont{K.}~\bibnamefont{Abe}} \bibnamefont{et~al.}
  (\bibinfo{collaboration}{SLD}), \bibinfo{journal}{Phys. Rev. Lett.}
  \textbf{\bibinfo{volume}{74}}, \bibinfo{pages}{1512} (\bibinfo{year}{1995}),
  \eprint{hep-ex/9501006}.

\bibitem[{\citenamefont{Barone and Ratcliffe}(2003)}]{Barone:2003fy}
\bibinfo{author}{\bibfnamefont{V.}~\bibnamefont{Barone}} \bibnamefont{and}
  \bibinfo{author}{\bibfnamefont{P.~G.} \bibnamefont{Ratcliffe}},
  \emph{\bibinfo{title}{Transverse Spin Physics}} (\bibinfo{publisher}{World
  Scientific}, \bibinfo{address}{River Edge, USA}, \bibinfo{year}{2003}).

\bibitem[{\citenamefont{Ralston and Soper}(1979)}]{Ralston:1979ys}
\bibinfo{author}{\bibfnamefont{J.~P.} \bibnamefont{Ralston}} \bibnamefont{and}
  \bibinfo{author}{\bibfnamefont{D.~E.} \bibnamefont{Soper}},
  \bibinfo{journal}{Nucl. Phys.} \textbf{\bibinfo{volume}{B152}},
  \bibinfo{pages}{109} (\bibinfo{year}{1979}).

\bibitem[{\citenamefont{Bianconi and Radici}(2006)}]{Bianconi:2005yj}
\bibinfo{author}{\bibfnamefont{A.}~\bibnamefont{Bianconi}} \bibnamefont{and}
  \bibinfo{author}{\bibfnamefont{M.}~\bibnamefont{Radici}},
  \bibinfo{journal}{Phys. Rev.} \textbf{\bibinfo{volume}{D73}},
  \bibinfo{pages}{034018} (\bibinfo{year}{2006}), \eprint{hep-ph/0512091}.

\bibitem[{\citenamefont{Soffer et~al.}(2002)\citenamefont{Soffer, Stratmann,
  and Vogelsang}}]{Soffer:2002tf}
\bibinfo{author}{\bibfnamefont{J.}~\bibnamefont{Soffer}},
  \bibinfo{author}{\bibfnamefont{M.}~\bibnamefont{Stratmann}},
  \bibnamefont{and}
  \bibinfo{author}{\bibfnamefont{W.}~\bibnamefont{Vogelsang}},
  \bibinfo{journal}{Phys. Rev.} \textbf{\bibinfo{volume}{D65}},
  \bibinfo{pages}{114024} (\bibinfo{year}{2002}), \eprint{hep-ph/0204058}.

\bibitem[{\citenamefont{Bianconi and
  Radici}(2005{\natexlab{a}})}]{Bianconi:2004wu}
\bibinfo{author}{\bibfnamefont{A.}~\bibnamefont{Bianconi}} \bibnamefont{and}
  \bibinfo{author}{\bibfnamefont{M.}~\bibnamefont{Radici}},
  \bibinfo{journal}{Phys. Rev.} \textbf{\bibinfo{volume}{D71}},
  \bibinfo{pages}{074014} (\bibinfo{year}{2005}{\natexlab{a}}),
  \eprint{hep-ph/0412368}.

\bibitem[{\citenamefont{Bianconi and
  Radici}(2005{\natexlab{b}})}]{Bianconi:2005bd}
\bibinfo{author}{\bibfnamefont{A.}~\bibnamefont{Bianconi}} \bibnamefont{and}
  \bibinfo{author}{\bibfnamefont{M.}~\bibnamefont{Radici}},
  \bibinfo{journal}{Phys. Rev.} \textbf{\bibinfo{volume}{D72}},
  \bibinfo{pages}{074013} (\bibinfo{year}{2005}{\natexlab{b}}),
  \eprint{hep-ph/0504261}.

\bibitem[{\citenamefont{Anselmino et~al.}(2003)\citenamefont{Anselmino,
  D'Alesio, and Murgia}}]{Anselmino:2002pd}
\bibinfo{author}{\bibfnamefont{M.}~\bibnamefont{Anselmino}},
  \bibinfo{author}{\bibfnamefont{U.}~\bibnamefont{D'Alesio}}, \bibnamefont{and}
  \bibinfo{author}{\bibfnamefont{F.}~\bibnamefont{Murgia}},
  \bibinfo{journal}{Phys. Rev.} \textbf{\bibinfo{volume}{D67}},
  \bibinfo{pages}{074010} (\bibinfo{year}{2003}), \eprint{hep-ph/0210371}.

\bibitem[{\citenamefont{Efremov et~al.}(2004)\citenamefont{Efremov, Goeke, and
  Schweitzer}}]{Efremov:2004qs}
\bibinfo{author}{\bibfnamefont{A.~V.} \bibnamefont{Efremov}},
  \bibinfo{author}{\bibfnamefont{K.}~\bibnamefont{Goeke}}, \bibnamefont{and}
  \bibinfo{author}{\bibfnamefont{P.}~\bibnamefont{Schweitzer}},
  \bibinfo{journal}{Eur. Phys. J.} \textbf{\bibinfo{volume}{C35}},
  \bibinfo{pages}{207} (\bibinfo{year}{2004}), \eprint{hep-ph/0403124}.

\bibitem[{\citenamefont{Airapetian et~al.}(2005)}]{Airapetian:2004tw}
\bibinfo{author}{\bibfnamefont{A.}~\bibnamefont{Airapetian}}
  \bibnamefont{et~al.} (\bibinfo{collaboration}{HERMES}),
  \bibinfo{journal}{Phys. Rev. Lett.} \textbf{\bibinfo{volume}{94}},
  \bibinfo{pages}{012002} (\bibinfo{year}{2005}), \eprint{hep-ex/0408013}.

\bibitem[{\citenamefont{Diefenthaler}(2005)}]{Diefenthaler:2005gx}
\bibinfo{author}{\bibfnamefont{M.}~\bibnamefont{Diefenthaler}},
  \bibinfo{journal}{AIP Conf. Proc.} \textbf{\bibinfo{volume}{792}},
  \bibinfo{pages}{933} (\bibinfo{year}{2005}), \eprint{hep-ex/0507013}.

\bibitem[{\citenamefont{Alexakhin et~al.}(2005)}]{Alexakhin:2005iw}
\bibinfo{author}{\bibfnamefont{V.~Y.} \bibnamefont{Alexakhin}}
  \bibnamefont{et~al.} (\bibinfo{collaboration}{COMPASS}),
  \bibinfo{journal}{Phys. Rev. Lett.} \textbf{\bibinfo{volume}{94}},
  \bibinfo{pages}{202002} (\bibinfo{year}{2005}), \eprint{hep-ex/0503002}.

\bibitem[{\citenamefont{Collins}(1993)}]{Collins:1993kk}
\bibinfo{author}{\bibfnamefont{J.~C.} \bibnamefont{Collins}},
  \bibinfo{journal}{Nucl. Phys.} \textbf{\bibinfo{volume}{B396}},
  \bibinfo{pages}{161} (\bibinfo{year}{1993}),
  \eprint[http://arXiv.org/abs]{hep-ph/9208213}.

\bibitem[{\citenamefont{Boer and Mulders}(1998)}]{Boer:1998nt}
\bibinfo{author}{\bibfnamefont{D.}~\bibnamefont{Boer}} \bibnamefont{and}
  \bibinfo{author}{\bibfnamefont{P.~J.} \bibnamefont{Mulders}},
  \bibinfo{journal}{Phys. Rev.} \textbf{\bibinfo{volume}{D57}},
  \bibinfo{pages}{5780} (\bibinfo{year}{1998}),
  \eprint[http://arXiv.org/abs]{hep-ph/9711485}.

\bibitem[{\citenamefont{Collins and Metz}(2004)}]{Collins:2004nx}
\bibinfo{author}{\bibfnamefont{J.~C.} \bibnamefont{Collins}} \bibnamefont{and}
  \bibinfo{author}{\bibfnamefont{A.}~\bibnamefont{Metz}},
  \bibinfo{journal}{Phys. Rev. Lett.} \textbf{\bibinfo{volume}{93}},
  \bibinfo{pages}{252001} (\bibinfo{year}{2004}), \eprint{hep-ph/0408249}.

\bibitem[{\citenamefont{Ji et~al.}(2005)\citenamefont{Ji, Ma, and
  Yuan}}]{Ji:2004wu}
\bibinfo{author}{\bibfnamefont{X.}~\bibnamefont{Ji}},
  \bibinfo{author}{\bibfnamefont{J.-P.} \bibnamefont{Ma}}, \bibnamefont{and}
  \bibinfo{author}{\bibfnamefont{F.}~\bibnamefont{Yuan}},
  \bibinfo{journal}{Phys. Rev.} \textbf{\bibinfo{volume}{D71}},
  \bibinfo{pages}{034005} (\bibinfo{year}{2005}), \eprint{hep-ph/0404183}.

\bibitem[{\citenamefont{Radici et~al.}(2002)\citenamefont{Radici, Jakob, and
  Bianconi}}]{Radici:2001na}
\bibinfo{author}{\bibfnamefont{M.}~\bibnamefont{Radici}},
  \bibinfo{author}{\bibfnamefont{R.}~\bibnamefont{Jakob}}, \bibnamefont{and}
  \bibinfo{author}{\bibfnamefont{A.}~\bibnamefont{Bianconi}},
  \bibinfo{journal}{Phys. Rev.} \textbf{\bibinfo{volume}{D65}},
  \bibinfo{pages}{074031} (\bibinfo{year}{2002}),
  \eprint[http://arXiv.org/abs]{hep-ph/0110252}.

\bibitem[{\citenamefont{van~der Nat}(2005)}]{vanderNat:2005yf}
\bibinfo{author}{\bibfnamefont{P.~B.} \bibnamefont{van~der Nat}}
  (\bibinfo{collaboration}{HERMES}) (\bibinfo{year}{2005}),
  \eprint{hep-ex/0512019}.

\bibitem[{\citenamefont{Martin}(2006)}]{Martin:2006}
\bibinfo{author}{\bibfnamefont{A.}~\bibnamefont{Martin}}
  (\bibinfo{collaboration}{COMPASS}) (\bibinfo{year}{2006}),
  \bibinfo{note}{talk presented at the Workshop on the QCD Structure of the
  Nucleon (QCD-N'06), Villa Mondragone, Rome, Italy, 12-16 June 2006.}

\bibitem[{\citenamefont{Abe et~al.}(2006)}]{Abe:2005zx}
\bibinfo{author}{\bibfnamefont{K.}~\bibnamefont{Abe}} \bibnamefont{et~al.}
  (\bibinfo{collaboration}{BELLE}), \bibinfo{journal}{Phys. Rev. Lett.}
  \textbf{\bibinfo{volume}{96}}, \bibinfo{pages}{232002}
  (\bibinfo{year}{2006}), \eprint{hep-ex/0507063}.

\bibitem[{\citenamefont{Hasuko et~al.}(2003)\citenamefont{Hasuko,
  Grosse~Perdekamp, Ogawa, Lange, and Siegle}}]{Hasuko:2003ay}
\bibinfo{author}{\bibfnamefont{K.}~\bibnamefont{Hasuko}},
  \bibinfo{author}{\bibfnamefont{M.}~\bibnamefont{Grosse~Perdekamp}},
  \bibinfo{author}{\bibfnamefont{A.}~\bibnamefont{Ogawa}},
  \bibinfo{author}{\bibfnamefont{J.~S.} \bibnamefont{Lange}}, \bibnamefont{and}
  \bibinfo{author}{\bibfnamefont{V.}~\bibnamefont{Siegle}},
  \bibinfo{journal}{AIP Conf. Proc.} \textbf{\bibinfo{volume}{675}},
  \bibinfo{pages}{454} (\bibinfo{year}{2003}).

\bibitem[{\citenamefont{Bianconi
  et~al.}(2000{\natexlab{b}})\citenamefont{Bianconi, Boffi, Jakob, and
  Radici}}]{Bianconi:1999uc}
\bibinfo{author}{\bibfnamefont{A.}~\bibnamefont{Bianconi}},
  \bibinfo{author}{\bibfnamefont{S.}~\bibnamefont{Boffi}},
  \bibinfo{author}{\bibfnamefont{R.}~\bibnamefont{Jakob}}, \bibnamefont{and}
  \bibinfo{author}{\bibfnamefont{M.}~\bibnamefont{Radici}},
  \bibinfo{journal}{Phys. Rev.} \textbf{\bibinfo{volume}{D62}},
  \bibinfo{pages}{034009} (\bibinfo{year}{2000}{\natexlab{b}}),
  \eprint[http://arXiv.org/abs]{hep-ph/9907488}.

\bibitem[{\citenamefont{Sjostrand et~al.}(2001)}]{Sjostrand:2000wi}
\bibinfo{author}{\bibfnamefont{T.}~\bibnamefont{Sjostrand}}
  \bibnamefont{et~al.}, \bibinfo{journal}{Comput. Phys. Commun.}
  \textbf{\bibinfo{volume}{135}}, \bibinfo{pages}{238} (\bibinfo{year}{2001}),
  \eprint{hep-ph/0010017}.

\bibitem[{\citenamefont{Liebing}(2004)}]{Liebing:2004us}
\bibinfo{author}{\bibfnamefont{P.}~\bibnamefont{Liebing}}, Ph.D. thesis,
  \bibinfo{school}{Universit{\"a}t Hamburg} (\bibinfo{year}{2004}),
  \bibinfo{note}{{DESY-THESIS-2004-036}}.

\bibitem[{\citenamefont{Bacchetta and Radici}(2003)}]{Bacchetta:2002ux}
\bibinfo{author}{\bibfnamefont{A.}~\bibnamefont{Bacchetta}} \bibnamefont{and}
  \bibinfo{author}{\bibfnamefont{M.}~\bibnamefont{Radici}},
  \bibinfo{journal}{Phys. Rev.} \textbf{\bibinfo{volume}{D67}},
  \bibinfo{pages}{094002} (\bibinfo{year}{2003}), \eprint{hep-ph/0212300}.

\bibitem[{\citenamefont{Boer et~al.}(2003)\citenamefont{Boer, Mulders, and
  Pijlman}}]{Boer:2003cm}
\bibinfo{author}{\bibfnamefont{D.}~\bibnamefont{Boer}},
  \bibinfo{author}{\bibfnamefont{P.~J.} \bibnamefont{Mulders}},
  \bibnamefont{and} \bibinfo{author}{\bibfnamefont{F.}~\bibnamefont{Pijlman}},
  \bibinfo{journal}{Nucl. Phys.} \textbf{\bibinfo{volume}{B667}},
  \bibinfo{pages}{201} (\bibinfo{year}{2003}), \eprint{hep-ph/0303034}.

\bibitem[{\citenamefont{Bacchetta and
  Radici}(2004{\natexlab{b}})}]{Bacchetta:2004mw}
\bibinfo{author}{\bibfnamefont{A.}~\bibnamefont{Bacchetta}} \bibnamefont{and}
  \bibinfo{author}{\bibfnamefont{M.}~\bibnamefont{Radici}}
  (\bibinfo{year}{2004}{\natexlab{b}}), \eprint{hep-ph/0412141}.

\bibitem[{\citenamefont{Bacchetta et~al.}(2004)\citenamefont{Bacchetta,
  D'Alesio, Diehl, and Miller}}]{Bacchetta:2004jz}
\bibinfo{author}{\bibfnamefont{A.}~\bibnamefont{Bacchetta}},
  \bibinfo{author}{\bibfnamefont{U.}~\bibnamefont{D'Alesio}},
  \bibinfo{author}{\bibfnamefont{M.}~\bibnamefont{Diehl}}, \bibnamefont{and}
  \bibinfo{author}{\bibfnamefont{C.~A.} \bibnamefont{Miller}},
  \bibinfo{journal}{Phys. Rev.} \textbf{\bibinfo{volume}{D70}},
  \bibinfo{pages}{117504} (\bibinfo{year}{2004}), \eprint{hep-ph/0410050}.

\bibitem[{\citenamefont{Caprini et~al.}(2006)\citenamefont{Caprini, Colangelo,
  and Leutwyler}}]{Caprini:2005zr}
\bibinfo{author}{\bibfnamefont{I.}~\bibnamefont{Caprini}},
  \bibinfo{author}{\bibfnamefont{G.}~\bibnamefont{Colangelo}},
  \bibnamefont{and}
  \bibinfo{author}{\bibfnamefont{H.}~\bibnamefont{Leutwyler}},
  \bibinfo{journal}{Phys. Rev. Lett.} \textbf{\bibinfo{volume}{96}},
  \bibinfo{pages}{132001} (\bibinfo{year}{2006}), \eprint{hep-ph/0512364}.

\bibitem[{\citenamefont{Kitagawa and Sakemi}(2000)}]{Kitagawa:2000ji}
\bibinfo{author}{\bibfnamefont{H.}~\bibnamefont{Kitagawa}} \bibnamefont{and}
  \bibinfo{author}{\bibfnamefont{Y.}~\bibnamefont{Sakemi}},
  \bibinfo{journal}{Prog. Theor. Phys.} \textbf{\bibinfo{volume}{104}},
  \bibinfo{pages}{421} (\bibinfo{year}{2000}).

\bibitem[{\citenamefont{Lafferty}(1993)}]{Lafferty:1993sx}
\bibinfo{author}{\bibfnamefont{G.~D.} \bibnamefont{Lafferty}},
  \bibinfo{journal}{Z. Phys.} \textbf{\bibinfo{volume}{C60}},
  \bibinfo{pages}{659} (\bibinfo{year}{1993}).

\bibitem[{\citenamefont{Gamberg et~al.}(2003)\citenamefont{Gamberg, Goldstein,
  and Oganessyan}}]{Gamberg:2003eg}
\bibinfo{author}{\bibfnamefont{L.~P.} \bibnamefont{Gamberg}},
  \bibinfo{author}{\bibfnamefont{G.~R.} \bibnamefont{Goldstein}},
  \bibnamefont{and} \bibinfo{author}{\bibfnamefont{K.~A.}
  \bibnamefont{Oganessyan}}, \bibinfo{journal}{Phys. Rev.}
  \textbf{\bibinfo{volume}{D68}}, \bibinfo{pages}{051501(R)}
  (\bibinfo{year}{2003}), \eprint{hep-ph/0307139}.

\bibitem[{\citenamefont{Bacchetta et~al.}(2002)\citenamefont{Bacchetta, Kundu,
  Metz, and Mulders}}]{Bacchetta:2002tk}
\bibinfo{author}{\bibfnamefont{A.}~\bibnamefont{Bacchetta}},
  \bibinfo{author}{\bibfnamefont{R.}~\bibnamefont{Kundu}},
  \bibinfo{author}{\bibfnamefont{A.}~\bibnamefont{Metz}}, \bibnamefont{and}
  \bibinfo{author}{\bibfnamefont{P.~J.} \bibnamefont{Mulders}},
  \bibinfo{journal}{Phys. Rev.} \textbf{\bibinfo{volume}{D65}},
  \bibinfo{pages}{094021} (\bibinfo{year}{2002}), \eprint{hep-ph/0201091}.

\bibitem[{\citenamefont{Amrath et~al.}(2005)\citenamefont{Amrath, Bacchetta,
  and Metz}}]{Amrath:2005gv}
\bibinfo{author}{\bibfnamefont{D.}~\bibnamefont{Amrath}},
  \bibinfo{author}{\bibfnamefont{A.}~\bibnamefont{Bacchetta}},
  \bibnamefont{and} \bibinfo{author}{\bibfnamefont{A.}~\bibnamefont{Metz}},
  \bibinfo{journal}{Phys. Rev.} \textbf{\bibinfo{volume}{D71}},
  \bibinfo{pages}{114018} (\bibinfo{year}{2005}), \eprint{hep-ph/0504124}.

\bibitem[{\citenamefont{Estabrooks and Martin}(1974)}]{Estabrooks:1974vu}
\bibinfo{author}{\bibfnamefont{P.}~\bibnamefont{Estabrooks}} \bibnamefont{and}
  \bibinfo{author}{\bibfnamefont{A.~D.} \bibnamefont{Martin}},
  \bibinfo{journal}{Nucl. Phys.} \textbf{\bibinfo{volume}{B79}},
  \bibinfo{pages}{301} (\bibinfo{year}{1974}).

\bibitem[{\citenamefont{Eidelman et~al.}(2004)}]{Eidelman:2004wy}
\bibinfo{author}{\bibfnamefont{S.}~\bibnamefont{Eidelman}} \bibnamefont{et~al.}
  (\bibinfo{collaboration}{Particle Data Group}), \bibinfo{journal}{Phys.
  Lett.} \textbf{\bibinfo{volume}{B592}}, \bibinfo{pages}{1}
  (\bibinfo{year}{2004}).

\bibitem[{\citenamefont{Colangelo et~al.}(2001)\citenamefont{Colangelo, Gasser,
  and Leutwyler}}]{Colangelo:2001df}
\bibinfo{author}{\bibfnamefont{G.}~\bibnamefont{Colangelo}},
  \bibinfo{author}{\bibfnamefont{J.}~\bibnamefont{Gasser}}, \bibnamefont{and}
  \bibinfo{author}{\bibfnamefont{H.}~\bibnamefont{Leutwyler}},
  \bibinfo{journal}{Nucl. Phys.} \textbf{\bibinfo{volume}{B603}},
  \bibinfo{pages}{125} (\bibinfo{year}{2001}), \eprint{hep-ph/0103088}.

\bibitem[{\citenamefont{Wakamatsu}(2001)}]{Wakamatsu:2000fd}
\bibinfo{author}{\bibfnamefont{M.}~\bibnamefont{Wakamatsu}},
  \bibinfo{journal}{Phys. Lett.} \textbf{\bibinfo{volume}{B509}},
  \bibinfo{pages}{59} (\bibinfo{year}{2001}), \eprint{hep-ph/0012331}.

\bibitem[{\citenamefont{Korotkov et~al.}(2001)\citenamefont{Korotkov, Nowak,
  and Oganessyan}}]{Korotkov:2001jx}
\bibinfo{author}{\bibfnamefont{V.~A.} \bibnamefont{Korotkov}},
  \bibinfo{author}{\bibfnamefont{W.~D.} \bibnamefont{Nowak}}, \bibnamefont{and}
  \bibinfo{author}{\bibfnamefont{K.~A.} \bibnamefont{Oganessyan}},
  \bibinfo{journal}{Eur. Phys. J.} \textbf{\bibinfo{volume}{C18}},
  \bibinfo{pages}{639} (\bibinfo{year}{2001}),
  \eprint[http://arXiv.org/abs]{hep-ph/0002268}.

\bibitem[{\citenamefont{Schweitzer et~al.}(2001)}]{Schweitzer:2001sr}
\bibinfo{author}{\bibfnamefont{P.}~\bibnamefont{Schweitzer}}
  \bibnamefont{et~al.}, \bibinfo{journal}{Phys. Rev.}
  \textbf{\bibinfo{volume}{D64}}, \bibinfo{pages}{034013}
  (\bibinfo{year}{2001}), \eprint{hep-ph/0101300}.

\bibitem[{\citenamefont{Gluck et~al.}(1998)\citenamefont{Gluck, Reya, and
  Vogt}}]{Gluck:1998xa}
\bibinfo{author}{\bibfnamefont{M.}~\bibnamefont{Gluck}},
  \bibinfo{author}{\bibfnamefont{E.}~\bibnamefont{Reya}}, \bibnamefont{and}
  \bibinfo{author}{\bibfnamefont{A.}~\bibnamefont{Vogt}},
  \bibinfo{journal}{Eur. Phys. J.} \textbf{\bibinfo{volume}{C5}},
  \bibinfo{pages}{461} (\bibinfo{year}{1998}), \eprint{hep-ph/9806404}.

\end{thebibliography}

\end{document}